\documentclass[12pt]{article} 
\usepackage[utf8x]{inputenc} 
\usepackage[T1]{fontenc} 
\usepackage{amsmath,amssymb,bbm,bm,commath,mathtools,slashed,wasysym,xfrac} 
\usepackage{sectsty} 
\usepackage{graphicx} 
\graphicspath{{figures/}}
\usepackage{color} 
\usepackage{cite} 

\usepackage{geometry} 
\usepackage{fancyhdr} 
\usepackage{setspace} 
\usepackage[nottoc,notlof,notlot]{tocbibind} 
\usepackage[titles]{tocloft} 

\usepackage[unicode]{hyperref} 
\hypersetup{bookmarksnumbered=true, bookmarksopen=true, bookmarksopenlevel=2, breaklinks=true, citecolor=blue, colorlinks=true, linkcolor=red, linktoc=page, pdfborder={0 0 0}, pdfstartview=FitH, pdfauthor={DJain}, pdfsubject={Free energy and twisted index of 3d N=2 CSm theories}, pdftitle={3d N=2 ADE CS Quivers}, plainpages=false, unicode=true, urlcolor=cyan}

\DeclareUnicodeCharacter{"0393}{\Gamma}
\DeclareUnicodeCharacter{"0394}{\Delta}
\DeclareUnicodeCharacter{"0398}{\Theta}
\DeclareUnicodeCharacter{"039B}{\Lambda}
\DeclareUnicodeCharacter{"039E}{\Xi}
\DeclareUnicodeCharacter{"03A0}{\Pi}
\DeclareUnicodeCharacter{"03A3}{\Sigma}
\DeclareUnicodeCharacter{"03A5}{\Upsilon}
\DeclareUnicodeCharacter{"03A6}{\Phi}
\DeclareUnicodeCharacter{"03A8}{\Psi}
\DeclareUnicodeCharacter{"03A9}{\Omega}
\DeclareUnicodeCharacter{"03B1}{\alpha}
\DeclareUnicodeCharacter{"03B2}{\beta}
\DeclareUnicodeCharacter{"03B3}{\gamma}
\DeclareUnicodeCharacter{"03B4}{\delta}
\DeclareUnicodeCharacter{"03B5}{\epsilon}
\DeclareUnicodeCharacter{"03B6}{\zeta}
\DeclareUnicodeCharacter{"03B7}{\eta}
\DeclareUnicodeCharacter{"03B8}{\theta}
\DeclareUnicodeCharacter{"03D1}{\vartheta}
\DeclareUnicodeCharacter{"03B9}{\iota}
\DeclareUnicodeCharacter{"03BA}{\kappa}
\DeclareUnicodeCharacter{"03BB}{\lambda}
\DeclareUnicodeCharacter{"03BC}{\mu}
\DeclareUnicodeCharacter{"03BD}{\nu}
\DeclareUnicodeCharacter{"03BE}{\xi}
\DeclareUnicodeCharacter{"03C0}{\pi}
\DeclareUnicodeCharacter{"03C1}{\rho}
\DeclareUnicodeCharacter{"03C3}{\sigma} 
\DeclareUnicodeCharacter{"03C4}{\tau}
\DeclareUnicodeCharacter{"03C5}{\upsilon}
\DeclareUnicodeCharacter{"03C6}{\phi}
\DeclareUnicodeCharacter{"03D5}{\varphi}
\DeclareUnicodeCharacter{"03C7}{\chi}
\DeclareUnicodeCharacter{"03C8}{\psi}
\DeclareUnicodeCharacter{"03C9}{\omega}
\DeclareUnicodeCharacter{"21D0}{\Leftarrow}
\DeclareUnicodeCharacter{"0212B}{\AA}
\DeclareUnicodeCharacter{"00B7}{\cdot}
\DeclareUnicodeCharacter{"266A}{\eighthnote}
\DeclareUnicodeCharacter{"266B}{\twonotes}
\PrerenderUnicode{Σ}\PrerenderUnicode{×}
\PrerenderUnicode{¹}\PrerenderUnicode{³}
\PrerenderUnicode{₃}\PrerenderUnicode{₄}

\definecolor{title}{rgb}{0.5,0.3,0.7}
\definecolor{abst}{rgb}{0.366,0.366,0.266}
\definecolor{sect}{rgb}{0.5,0.1,0.5}
\definecolor{ssect}{rgb}{0.5,0.05,0.25}
\definecolor{sssect}{rgb}{0.5,0.025,0.125}
\definecolor{appsect}{rgb}{0.5,0.1,0.5}
\definecolor{ref}{rgb}{0.5,0.3,0.5}
\newcommand{\Title}[1] {\title{\color{title}\Huge #1}}
\newcommand{\TPheader}[3] {\date{}\maketitle\thispagestyle{fancy}\pagenumbering{alph}\lhead{#1}\chead{#2}\rhead{#3}\cfoot{}}
\newcommand{\makepage}[1] {\newpage\pagenumbering{#1}}

\newcommand{\Abstract}[1] {\begin{abstract}\normalsize #1 \end{abstract}}

\sectionfont{\color{sect}\Large\bf}
\subsectionfont{\color{ssect}\large\bf}
\subsubsectionfont{\color{sssect}\normalsize\bf}
\renewcommand{\appendix}{\setcounter{section}{0}\sectionfont{\color{appsect}\Large\bf}\renewcommand{\thesection}{\Alph{section}}\renewcommand*{\theHsection}{app.\the\value{section}}} 
\newcommand\references[1]{\sectionfont{\color{ref}\Large\bf}\bibliographystyle{hephys}\bibliography{#1}}

\newcommand\eqs[1] {\begin{align}#1\end{align}}
\newcommand\eqsn[1] {\begin{align*}#1\end{align*}}

\newcommand\eqst[1] {\begin{multline}#1\end{multline}}
\newcommand\eqstn[1] {\begin{multline*}#1\end{multline*}}
\newcommand\eqsc[1] {\begin{gather}#1\end{gather}}
\newcommand\eqscn[1] {\begin{gather*}#1\end{gather*}}
\newcommand\equ[1] {\begin{equation}#1\end{equation}}
\newcommand\equa[1] {\equ{\begin{aligned}#1\end{aligned}}}
\newcommand\equn[1] {\begin{equation*}#1\end{equation*}}
\newcommand\equg[1] {\equ{\begin{gathered}#1\end{gathered}}}
\newcommand\fig[2] {\begin{figure}[#1]\centering #2\end{figure}}
\newcommand\pmat[1] {\begin{pmatrix}#1\end{pmatrix}}
\newcommand\spmat[1] {\big(\begin{smallmatrix}#1\end{smallmatrix}\big)}
\newcommand\tabl[2] {\begin{table}[#1]\centering #2\end{table}}

\renewcommand\exp[1] {e^{#1}}
\renewcommand\i {\dot{\iota}}
\newcommand\half {\tfrac{1}{2}}
\newcommand\s {\sigma}

\renewcommand\( {\left(}
\renewcommand\) {\right)}
\newcommand\wh {\widehat}

\DeclareMathOperator{\Li}{Li}

\DeclareMathOperator{\sgn}{sgn}
\DeclareMathOperator{\tr}{tr}
\DeclareMathOperator{\Ar}{Area}
\DeclareMathOperator{\Vol}{Vol}

\newcommand\B {{\cal B}}

\newcommand\I {{\cal I}}

\newcommand\M {{\cal M}}
\newcommand\N {{\cal N}} 
\renewcommand\O {{\cal O}}
\renewcommand\P {{\cal P}}
\newcommand\V {{\cal V}}
\newcommand\W {{\cal W}}
\newcommand\bC {{\mathbb C}}

\newcommand\bR {{\mathbb R}}

\newcommand\bZ {{\mathbb Z}}
\newcommand\fg {\mathfrak{g}}
\newcommand\fm {\mathfrak{m}}
\newcommand\fn {\mathfrak{n}}

\newcommand\nn {\nonumber\\}
\newcommand\VY {\Vol(Y_7)}
\newcommand\Vb {\bar{\V}}
\newcommand\Ib {\bar{\I}}

\numberwithin{equation}{section} 
\begin{document}
\Title{\texorpdfstring{3d $\N=2$ $\wh{ADE}$ Chern-Simons Quivers}{3d N=2 Affine-ADE Chern-Simons Quivers}}

\author{Dharmesh Jain\footnote{\href{mailto:d.jain@saha.ac.in}{d.jain@saha.ac.in}}\; and Augniva Ray\footnote{\href{mailto:augniva.ray@saha.ac.in}{augniva.ray@saha.ac.in}}\bigskip\\
\emph{Theory Division, Saha Institute of Nuclear Physics, HBNI,}\\ \emph{1/AF Bidhan Nagar, Kolkata 700064, India}
}

\TPheader{}{\today}{} 

\Abstract{We study 3d $\N=2$ Chern-Simons (CS) quiver theories on $S^3$ and $Σ_{\fg}×S^1$. Using localization results, we examine their partition functions in the large rank limit and requiring the resulting matrix models to be local, find a large class of quiver theories that include quivers in one-to-one correspondence with the $\wh{ADE}$ Dynkin diagrams. We compute explicitly the partition function on $S^3$ for $\wh{D}$ quivers and that on $Σ_{\fg}×S^1$ for $\wh{AD}$ quivers, which lead to certain predictions for their holographic duals. We also provide a new and simple proof of the ``index theorem'', extending its applicability to a larger class of theories than considered before in the literature.
}

\makepage{Roman} 
\tableofcontents
\makepage{arabic}

\section{Introduction and Outline}\label{sec:IO}
Supersymmetric localization has made a whole host of theories accessible to non-perturbative analysis. It provides a powerful framework to construct and compute quantities along the renormalization group flow non-perturbatively. One of them is the exact partition function $Z$ for supersymmetric gauge theories put on various curved manifolds in different dimensions (see \cite{Pestun:2016zxk} and references therein). This has led to a deeper understanding, checks and / or discovery of various dualities among field theories, even across dimensions. Since one gets access to exact results, one can test AdS/CFT correspondence in the regime relating weak gravity results (may or may not have been obtained via localization) to strong coupling results in the field theory (highly likely to have been obtained via localization). We will focus here on this latter possibility with the study of supersymmetric quiver gauge theories in three-dimensions, i.e., an example of AdS${}_4$/CFT${}_3$.

\paragraph{Free Energy.} Localization was successfully applied to compute the partition function of 3d Chern-Simons-matter (CSm) theories placed on 3-sphere $S^3$ in \cite{Kapustin:2009kz,Jafferis:2010un,Hama:2010av}. The first explicit construction of a 3d $\N=6$ gauge theory with M-theory dual was presented in \cite{Aharony:2008ug} and is now known as ABJM theory. It involves two $U(N)$ gauge groups with CS terms at levels $±k$ and four bifundamental chiral multiplets (in terms of $\N=2$ multiplets). The dual geometry involved placing $N$ M2-branes at the tip of a $\bC^4/\bZ_k$ singularity such that in large $N$ limit, the AdS${}_4\times S^7/\bZ_k$ vacuum solution of M-theory was obtained. Following this, a large number of $\N≥2$ dual pairs have been identified, with the M-theory dual of the form AdS$_4\times Y_7$, where $Y_7$ is a (tri-)Sasaki-Einstein manifold given by the base of a certain 8d (hyper)kähler cone\cite{Jafferis:2008qz,Herzog:2010hf,Martelli:2011qj,Cheon:2011vi,Jafferis:2011zi,Gulotta:2011si,Amariti:2011uw}. The AdS/CFT dictionary relates $\VY$ to the free energy $F_{S^{3}}$ of the dual gauge theories in the large $N$ limit\cite{Drukker:2010nc,Herzog:2010hf}
\equ{F_{S^3}=-\log|Z_{S^3}|=N^{\sfrac{3}{2}}\sqrt{\frac{2\pi^{6}}{27\VY}}\,·
\label{FS3vol}}
This provides an important tool to compute the volumes via computations in the dual field theory.

We will consider general $\N=2$ quiver gauge theories on $S^3$ involving matter multiplets with arbitrary R-charges $Δ$'s in the large $N$ limit. This will lead us to a large class of quiver theories whose free energy scales as $N^{\sfrac{3}{2}}$ from requiring that the long-range forces in the resulting matrix model cancel (or equivalently, that the matrix model be local) along with a constraint on the R-charges of bifundamental multiplets given by \eqref{ADEconditionFS3}. A subset of this constraint leads to the $\wh{ADE}$ classification via a simple constraint on the bifundamental R-charges:
\equ{Δ_{(a,b)}+Δ_{(b,a)}=1\,.
}
Note that for $\N≥3$ case, this condition is automatic since the supersymmetry enhancement fixes the R-charges to be $\half$ and $\wh{ADE}$ classification was presented in \cite{Gulotta:2011vp}. We will then explicitly solve the large $N$ matrix model of the $\N=2$ $\wh{D}$ quiver theories\footnote{The $\N=2$ $\wh{A}$ quivers have been discussed in detail in \cite{Gulotta:2011aa} and $\wh{E}$ quivers can be solved using the approach discussed in this paper, but due to increasing complexity (and decreasing clarity) of the expressions, we refrain from giving the explicit results here.}, whose dual geometry involves certain 7-dimensional Sasaki-Einstein manifolds $Y_7$. Computation of their volumes directly does not necessarily give the volume for the Calabi-Yau (CY) metric necessary for the AdS/CFT correspondence\footnote{It was not the case for $\N≥3$ theories where the hyperkähler structure guarantees the CY condition, which was used to calculate explicit volumes for toric quivers like $\wh{A}$ in \cite{Yee:2006ba} and nontoric ones like $\wh{D}$ in \cite{Crichigno:2017rqg}.}. This can be circumvented by using the geometrical result of volume minimization that fixes the Reeb vector and gives the correct volume of the Ricci-flat Kähler manifold, which corresponds in the dual field theory to $F$-maximization that fixes the R-charges at the IR fixed point \cite{Martelli:2005tp,Martelli:2006yb,Martelli:2011qj}. We will leave the check of this correspondence in the case of $\wh{D}$ quivers for future work and treat the $F_{S^3}$ computed in Section \ref{sec:FEV} as predicting the volumes of the relevant Sasaki-Einstein $Y_7$'s.

\paragraph{Twisted Index.} Localization has also been used to compute the partition function of 3d CSm theories on $Σ_{\fg}×S^1$ with a partial topological twist \cite{Witten:1988ze} on the Riemann surface ($Σ_{\fg}$) of genus $\fg$\cite{Benini:2015noa,Benini:2016hjo,Closset:2016arn}. This partition function is usually called topologically twisted index and depends on chemical potentials $ν≡A_t^{bg}+\i\s^{bg}$ (complex mass parameters constructed from the background vector multiplets coupled to the flavour symmetries) as well as background magnetic fluxes $\fn$ through $Σ_{\fg}$ for the flavour and R-symmetry. It was shown in \cite{Benini:2015eyy} that the large $N$ limit of $\Re \log Z_{S^2×S^1}$ for ABJM theory reproduces the macroscopic entropy $S_{BH}$ of supersymmetric magnetic AdS${}_4$ black holes discussed in \cite{Cacciatori:2009iz}. The large $N$ limit for many other theories has been considered in \cite{Hosseini:2016tor,Hosseini:2016ume}, which revealed a connection of Bethe potential $\V$ -- obtained as an intermediate step while computing the twisted index -- to the $F_{S^3}$ discussed above. In addition, a relation dubbed ``index theorem''\footnote{This nomenclature was introduced in \cite{Hosseini:2016tor} and has stuck in the derivative literature since then. It has no relation to the (Atiyah-Singer) theorem about the index of elliptic differential operators.} was proven which showed that the twisted index could be written directly in terms of the $\V$ and its derivatives with respect to the chemical potentials.

We will again consider general $\N=2$ quiver gauge theories on $Σ_{\fg}×S^1$ in large $N$ limit and find that $\wh{ADE}$ classification [as a subset of quiver theories which satisfy \eqref{ADEconditionTTI} and \eqref{FFxcondition}] follows from the requirement that the matrix model is local and the following set of constraints is satisfied:
\equ{ν_{(a,b)}+ν_{(b,a)}=\half\qquad \text{and}\qquad \fn_{(a,b)}+\fn_{(b,a)}=1\,.
}
We will then compute the large $N$ limit of the topologically twisted index for $\wh{AD}$ quivers. Abusing the terminology slightly, we will denote $\I=\log|Z_{Σ_{\fg}×S^1}|$ and call it the twisted index most of the time\footnote{We will consider here field theories having M-theory duals only. Theories with type IIA duals can also be similarly considered as have been done in \cite{Benini:2016rke,Hosseini:2017fjo,Benini:2017oxt}.}. Along the way, we will extend (and simplify) the proof of the relation between the Bethe potential and the twisted index to cover not just the $\wh{A}$-type quiver theories\cite{Hosseini:2016tor} but a large class of theories including the $\wh{DE}$ quivers. Once again, we will not construct the dual AdS${}_4$ black hole solutions to compute the entropy $S_{BH}$ explicitly (see the recent review \cite{Zaffaroni:2019dhb} and references therein for more on twisted index and entropy matching). Assuming AdS/CFT correspondence to hold, we can conjecture that the twisted index computed in Section \ref{sec:TIE} for $\wh{AD}$ quivers is the entropy for the corresponding dual black holes (after extremization with respect to the chemical potentials), leaving an explicit check for future. However, for the specific case of the universal twist \cite{Bobev:2017uzs,Azzurli:2017kxo}, we provide further evidence for the AdS/CFT correspondence. In this case, due to holographic RG flow from AdS${}_4$ to AdS${}_2$, the black hole entropy follows a simple relation ($\fg>1$):
\equ{S_{BH}=(\fg-1)F_{S^3}\,.
}
The twisted index is also proportional to the free energy and a simple relation between various quantities introduced till now follows
\equ{S_{BH}[\tfrac{Δ}{2}]=\I[\tfrac{Δ}{2}]=(\fg-1)\bigg[4\V[\tfrac{Δ}{2}]=F_{S^3}[Δ]=\frac{4πN^{\sfrac{3}{2}}}{3}μ[Δ]\bigg]\quad \text{ with }\quad \frac{1}{8μ^2}=\frac{\VY}{\Vol(S^7)}\,·
\label{univTrel}}
Here, $Δ$'s are the R-charges of the bifundamental fields appearing in the $\wh{ADE}$ quiver at a superconformal fixed point where $F_{S^3}$ is extremized, i.e., $\frac{∂F_{S^3}}{∂Δ_{(a,b)}}=0$.

\paragraph{Outline.} In Section \ref{sec:S3} we review the computation of free energy on $S^3$ in large $N$ limit. In Section \ref{sec:TI} we revisit the twisted index computation in large $N$ limit and set up our notation consistent with the previous section. Along the way, we provide some new results including a simple proof of the relation between $\I$ and $\V$. In Section \ref{sec:FEV} we specialize to the free energy computation: we review the result for $\wh{A}_m$ quivers; provide an explicit example of $\wh{D}_4$ quiver and conjecture the result for $\wh{D}_n$ quivers. In Section \ref{sec:TIE} we move on to the twisted index computation: we provide explicit computations for $\wh{A}_3$ and $\wh{D}_4$ quivers, and present the general results for $\wh{A}_m$\footnote{To our knowledge, the general result for twisted index of $\wh{A}_m$ quivers presented here is new and only certain limits of that result are available in the literature.} and $\wh{D}_n$ quivers based on the previous section. We conclude with Section \ref{sec:SO} and the Appendix \ref{app:DP} where we collect some derivations and proofs to make this paper self-contained.

\section[\texorpdfstring{$S^3$ and Free Energy}{S³ and Free Energy}]{$\bm{S^3}$ and Free Energy}\label{sec:S3}
We consider $\N=2$ quiver CS gauge theories involving vector multiplets (VM) with gauge group $G=⊗_aU(N_a)$ and matter multiplets (MM) in representation $⊗_iR_i$ of $G$. We will deal with (anti-)bifundamental and (anti-)fundamental representations only. VM consists of a gauge field $A_μ$, an auxiliary complex fermion $λ_α$ ($α=1,2$) and two auxiliary real scalars $\s$ and $D$. MM consists of a complex scalar $φ$, a complex fermion $ψ_α$ and an auxiliary complex scalar $F$.

The theories in consideration have been localized on $S^3$ in \cite{Kapustin:2009kz,Jafferis:2010un,Hama:2010av}. According to them, $Z_{S^3}$ gets localized on configurations where $\s_a$ in the $\N=2$ VMs are constant $N_a×N_a$ matrices and thus the original path integral reduces to a matrix model:
\eqsc{Z_{S^3}=\frac{1}{|\W|}∫\bigg(∏_a∏_{\mathclap{\text{Cartan}}}d\s_a\bigg)\exp{\i π∑_a k_a\tr(\s_a^2)} ∏_a\det_{Ad}\(2\sinh(πα(\s_a))\) ∏_{\substack{\text{MM in}\\ \text{rep }R_i}}\det_{R_i}\big(e^{\ell(1-Δ_i+\iρ_i(\s))}\big)\,,\\
\ell(z)=\tfrac{\i}{2π}\Li_2\big(e^{2\pi\i z}\big)+\tfrac{\i π}{2}z^2 -z\log\big(1-e^{2\pi\i z}\big) -\tfrac{\i\pi}{12}\,;\qquad \ell'(z)=-πz\cot(πz)\,,
\label{defell}}
where $k_a$ are the CS levels of the VM corresponding to $U(N_a)$, $Δ_i$ are the R-charges of the corresponding MM in representation $R_i$, and $α(\s)$, $ρ(\s)$ are the roots and weights of the appropriate gauge group representations. Denoting the eigenvalues of $\s_a$ matrices by $λ_{a,i}$ with $i=1,⋯,N_a$ leads to a simple expression for free energy:
\begingroup
\allowdisplaybreaks
\eqs{F_{S^3} &=-\log |Z_{S^3}| ⇒ Z_{S^3}=∫∏_{a,i}dλ_{a,i} \exp{-F_{S^3}(\{λ_{a,i}\})} \nn[5mm]
⇒ F_{S^3} &≈ -\i π ∑_{a,i} k_aλ_{a,i}^2 -2∑_a∑_{i>j}\log\big|2\sinh\big(πλ_{a,i}-πλ_{a,j}\big)\big| \nn
&\quad -∑_{(a,b)∈E}∑_{i,j}\ell\big(1-Δ_{(a,b)}+\i(λ_{a,i}-λ_{b,j})\big) -∑_a∑_{\{f^a\}}∑_i\ell\big(1-Δ_{f^a} +\i λ_{a,i}\big).
\label{Fgenexp}}
\endgroup
Here, we have included only bifundamental and fundamental representations explicitly; the (anti-)reps can be similarly added and will be added below as required. We are most concerned with the above expression's large rank limit, keeping the CS levels fixed. For that purpose, we rewrite $N_a→n_aN$ for some integers $n_a(≥1)$ and then take $N→∞$ by going to a continuum limit. We will mostly follow \cite{Herzog:2010hf,Jafferis:2011zi,Gulotta:2011aa,Gulotta:2011vp,Crichigno:2012sk} in our saddle point analysis of $F_{S^3}$ so most of this section has appeared before in the literature in one form or the other, apart from the explicit identification of $\N=2$ $\wh{ADE}$ quivers.

The saddle point equation following from \eqref{Fgenexp} for $λ_{a,i}$ is:
\eqst{0=\frac{∂F_{S^3}}{∂λ_{a,i}} \propto 2∑_{j≠i}\coth[π(λ_{a,i}-λ_{a,j})] -∑_{b|(a,b)∈E,j}(1-Δ_{(a,b)}+\i(λ_{a,i}-λ_{b,j}))\coth[π(λ_{a,i}-λ_{b,j}+\i Δ_{(a,b)})] \\
-∑_{b|(a,b)∈E,j}(1-Δ_{(b,a)}-\i(λ_{a,i}-λ_{b,j}))\coth[π(λ_{a,i}-λ_{b,j} -\i Δ_{(b,a)})].
}
The CS term and terms from fundamental matter are subleading compared to the vector and bifundamental contribution so we do not write them above. To take the continuum limit, we assume the eigenvalue distribution for $U(n_aN)$ to be
\equ{λ_{a,i}→λ_{a,I}(x)=N^αx+\i y_{a,I}(x) \quad \text{ (with $I=1,⋯,n_a$),}
}
and introduce an eigenvalue density $ρ(x)=\frac{1}{N}∑_iδ(x-x_i)$ such that $∫dx ρ(x)=1$. This allows us to use the large argument approximation for $\coth[π(λ_{a,i}-λ_{b,j})]≈\sgn(x-x')$ and convert $∑_i → N∫dxρ(x)∑_I$. Note that if we demand the same number of bifundamental and anti-bifundamental matters at each edge, then no contributions arise at $\O(N^{1+α})$. The contribution at $\O(N)$ then gives a constraint on $n_a$'s and R-charges as follows:
\eqsc{0=\frac{∂F_{S^3}}{∂λ_{a,i}} \propto \bigg(2n_a -∑_{b|(a,b)∈E}(2-Δ_{(a,b)}-Δ_{(b,a)})n_b\bigg)N∫dx'ρ(x')\sgn(x-x') \nn
⇒2n_a =∑_{b|(a,b)∈E}(2-Δ_{(a,b)}-Δ_{(b,a)})n_b\,.
\label{ADEconditionFS3}}
This constraint originating from the saddle point analysis guarantees the cancellation of long-range forces and the expression for free energy will turn out to be local. We will present the off-shell expression for free energy with generic R-charges but for explicit computation of free energy, we will consider a stricter constraint: $Δ_{(a,b)}+Δ_{(b,a)}=1$. It is easy to see that this gives us an $\wh{ADE}$ classification (see Figure \ref{Fig:quivers}) for these $\N=2$ quivers just like in the $\N=3$ case\footnote{\label{ADEfootnote}This is not the only simple solution of \eqref{ADEconditionFS3}. For example, ABJM theory ($\wh{A}_1$) and other odd $\wh{A}$ quivers can still be constructed with the less strict condition: $Δ_{(a-1,a)}+Δ_{(a,a-1)}+Δ_{(a,a+1)}+Δ_{(a+1,a)}=2$. It would be interesting to study generic non-$\wh{ADE}$ theories with non-trivial constraints on $R$-charges compatible with \eqref{ADEconditionFS3}.}. This condition can also be motivated from the analysis of superpotential as discussed in \cite{Gulotta:2011aa}.
\fig{h!}{\includegraphics[scale=1]{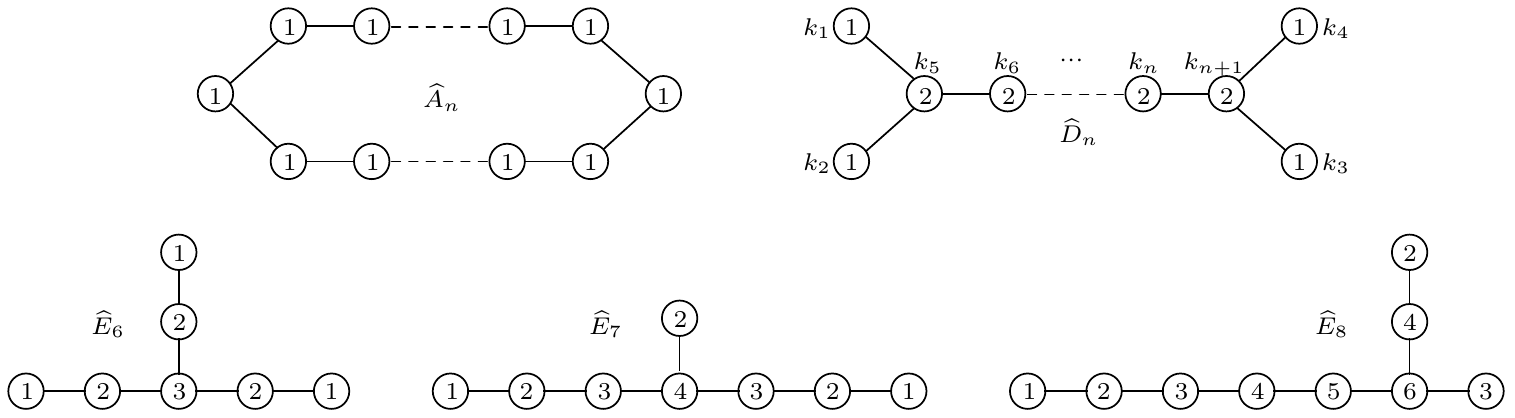}
\caption{$\wh{ADE}$ quivers with comarks $n_a$ written inside the nodes. (For $\wh{D}_n$ quivers, CS levels are also marked.)}
\label{Fig:quivers}}

Now moving to $F_{S^3}$, we have from \eqref{Fgenexp}
\eqsn{F_{S^3} &≈ -\i πN∫dx ρ(x)∑_{a,I}k_a \big(N^αx +\i y_{a,I}(x)\big)^2 \\
&-N^2∫dxdx'ρ(x)ρ(x')∑_{a,I,J}\log\left|2\sinh\big(πN^α(x-x')+\i π(y_{a,I}(x) -y_{a,J}(x'))\big)\right| \\
&-N^2∫dxdx'ρ(x)ρ(x')∑_{(a,b)∈E}∑_{I,J}\ell\big(1-Δ_{(a,b)} +\i N^α(x-x') -(y_{a,I}(x) -y_{b,J}(x'))\big) \\
&-N∫dxρ(x)∑_{\mathclap{a,\{f^a\},I}}\ell\big(1-Δ_{f^a} +\i N^αx -y_{a,I}\big)-N∫dxρ(x)∑_{\mathclap{a,\{\bar{f}^a\},I}}\ell\big(1-\bar{Δ}_{f^a} -\i N^αx +y_{a,I}\big).
}
We change variables from $N^α(x-x') → ξ$ where required and keep at most two highest orders of $N$ in each term to get\footnote{We use $\arg\(e^{2π\i x}\)=2πx+2π \left\lfloor\half-x\right\rfloor$ later. We have omitted divergent (as well as constant) terms that cancel due to \eqref{ADEconditionFS3}. See Appendix \ref{app:DP} for details.}
\eqs{F_{S^3} &≈ -\i πN^{1+2α}∑_a(n_ak_a)∫dx ρ(x)x^2 +2πN^{1+α}∫dx ρ(x)∑_{a,I}k_a x y_{a,I}(x) \nn
&+\frac{1}{4π}N^{2-α}∫dxρ(x)^2∑_{a,I,J}\arg\big(\exp{2π\i (y_{a,I}-y_{a,J}-\sfrac{1}{2})}\big)^2 \nn
&-\frac{1}{4π}N^{2-α}∫dxρ(x)^2∑_{(a,b)∈E}∑_{I,J}\Big[\big(1-Δ_{(a,b)}-(y_{a,I}-y_{b,J})\big)\arg\big(\exp{2π\i(\sfrac{1}{2}-Δ_{(a,b)}-(y_{a,I}-y_{b,J}))}\big)^2 \nn
&+\frac{1}{3π}\arg\big(\exp{2π\i(\sfrac{1}{2}-Δ_{(a,b)}-(y_{a,I}-y_{b,J}))}\big)\Big(π^2 -\arg\big(\exp{2π\i(\sfrac{1}{2}-Δ_{(a,b)}-(y_{a,I}-y_{b,J}))}\big)^2\Big) +(Δ_{(b,a)}\text{ terms})\Big] \nn
&+\frac{\i π}{2}N^{1+2α}∑_a(f^an_a -\bar{f}^an_a)∫dxρ(x)x^2 \nn
&+π N^{1+α}∑_{a,I}∫dxρ(x)|x|\bigg[∑_{\{f^a\}}\big(1-Δ_{f^a} -y_{a,I}\big) +∑_{\{\bar{f}^a\}}\big(1-\bar{Δ}_{f^a} +y_{a,I}\big)\bigg].
\label{FS3fullint}}
Here, $f^a(\bar{f}^a)$ are the total number of (anti-)fundamental fields at node $a$. We see 3 powers of $N$ so let us assume $∑_an_ak_a=0$ and $∑_a\(f^a -\bar{f}^a\)n_a=0$ so that we can match $N^{1+α}=N^{2-α}$ giving us the expected $α=\half$. We also point out that to get non-trivial solutions, a much stricter equality $f^a=\bar{f}^a$ needs to be imposed\footnote{While solving the matrix models explicitly, we will set $f^a=0$ since non-zero $f^a$ modify the resulting expressions in a well-known (and trivial) way (see for example \cite{Crichigno:2012sk,Jain:2015kzj}).} leading us to the final expression to be extremized:
\eqs{F_{S^3} &= N^{\sfrac{3}{2}}∫dx ρ(x)\Bigg[2πx ∑_{a,I}k_a y_{a,I}(x) +\frac{1}{4π}ρ(x)\bigg(∑_{a,I,J}\arg\big(\exp{2π\i (y_{a,I}-y_{a,J}-\sfrac{1}{2})}\big)^2 \nn
&-∑_{(a,b)∈E}∑_{I,J}\Big[\big(1-Δ_{(a,b)}-(y_{a,I}-y_{b,J})\big)\arg\big(\exp{2π\i(\sfrac{1}{2}-Δ_{(a,b)}-(y_{a,I}-y_{b,J}))}\big)^2 \nn
&+\frac{1}{3π}\arg\big(\exp{2π\i(\sfrac{1}{2}-Δ_{(a,b)}-(y_{a,I}-y_{b,J}))}\big)\Big(π^2 -\arg\big(\exp{2π\i(\sfrac{1}{2}-Δ_{(a,b)}-(y_{a,I}-y_{b,J}))}\big)^2\Big) +(Δ_{(b,a)}\text{ terms})\Big] \bigg) \nn
&+π|x|\(2n_F-Δ_F\)\Bigg] -2πμN^{\sfrac{3}{2}}\bigg(∫dx\,ρ(x) -1\bigg).
\label{FS3extfin}}
We defined $n_F=∑_af^an_a=∑_a\bar{f}^an_a$, $Δ_F=∑_a∑_{\{f^a\}}n_a\(Δ_{f^a}+\bar{Δ}_{f^a}\)$ and have added a Lagrange multiplier ($μ$) term to enforce the normalizability of the eigenvalue density. On general grounds\cite{Gulotta:2011si}, extremizing $F_{S^3}$ gives
\equ{\bar{F}_{S^3}=\frac{4πN^{\sfrac{3}{2}}}{3}μ.
\label{genextF}}
We will sometimes use a bar to denote an on-shell quantity as in \eqref{genextF} above, when compared to the off-shell quantity given by an integral expression as in \eqref{FS3extfin}.\\

\noindent This completes the review of the free energy $F_{S^3}$. Let us now turn to computation of the twisted index.

\section[\texorpdfstring{$Σ_{\fg}×S^1$ and Twisted Index}{Σg×S¹ and Twisted Index}]{$\bm{Σ_{\fg}×S^1}$ and Twisted Index}\label{sec:TI}
The topologically twisted index is the $Σ_{\fg}×S^1$ partition function with a topological twist along the Riemann surface of genus $\fg$, $Σ_{\fg}$. It was derived for $Σ_{\fg}=S^2$ in \cite{Benini:2015noa} and extended to generic $\fg$ in \cite{Benini:2016hjo}. The main result reads (we choose unit radius for the circle $S^1$):
\eqst{Z_{Σ_{\fg}×S^1}=\frac{1}{|\W|}∑_{\fm_a}∮\bigg(∏_a∏_{\mathclap{\text{Cartan}}}du_a\bigg)\B^{\fg}\,\exp{2π∑_a k_au_a·\fm_a} ∏_a\bigg(∏_{α∈G}\(1-e^{2πα(u_a)}\)^{1-\fg}∏_{α>0}(-1)^{α(\fm_a)}\bigg) \\
×∏_{I}∏_{ρ∈R_I}\bigg(\frac{e^{πρ(u_I)+π\i ν_I}}{1-e^{2πρ(u_I)+2π\i ν_I}}\bigg)^{ρ_I(\fm)+(\fg-1)\(\fn_I+(Δ_I-1)\)}\,,
\label{ZTIgenexp}}
where $u=\i\(∫_{S^1}A+\i\s\)$ are the holonomies and $\fm=\frac{1}{2π}∫_{Σ_{\fg}}F$ are the magnetic fluxes corresponding to the gauge group\footnote{We have kept the $(-1)^{α(\fm)}$ contribution of the vector multiplet explicitly as it contributes to the Bethe potential and is required for a consistent result, the way we take the large $N$ limit (see Appendix \ref{app:DP}).}, $ν=\(∫_{S^1}A^{bg}+\i\s^{bg}\)$ are the holonomies (or chemical potentials) and $\fn$ are the fluxes for the background vector multiplet coupled to flavour symmetry such that $\fn(\fg-1)$ is integer-quantized.\footnote{The different definitions for the same quantities corresponding to gauge and flavour groups are chosen for later convenience when comparing the large $N$ results for twisted index to those for free energy on $S^3$.} The real part of $ν$ is defined modulo 1 so we choose $ν$ to satisfy $0<ν<1$. The Hessian $\B$ is a contribution due to fermionic zero-modes and (up to some constant factors) is given by $\B≈\det_{ai,bj}\frac{∂^2Z_{\text{cl+1-loop}}}{∂u_a^i∂\fm_b^j}$, where $Z_{\text{cl+1-loop}}$ is the full integrand appearing in \eqref{ZTIgenexp} except for the $\B^{\fg}$ factor.

As with $Z_{S^3}$, we are interested in the large $N$ limit of the above expression. It was studied for ABJM theory $(\wh{A}_1)$ in \cite{Benini:2015eyy} and for $\wh{A}$-type quiver theories in \cite{Hosseini:2016tor}. As discussed there, due to the sum over magnetic fluxes, evaluating this limit becomes a two-step process: (1) Sum over magnetic fluxes, $\fm_a$; (2) Integrate over the holonomies, $u_a$. The first step involves summing a geometric series, which generates factors like $\frac{1}{1-e^{\i\B_a^i(u_a)}}$ leading to poles at $\hat{u}_a$ such that $e^{\i\B_a^i(\hat{u}_a)}=1$. We solve for $\hat{u}_a$ by constructing an auxiliary object called the ``Bethe potential'' $\V$ defined as $\frac{∂\V}{∂u_a^i}=\B_a^i$ such that extremizing $\V$ gives the Bethe ansatz equations (BAEs): $\frac{∂\V}{∂u_a^i}\big|_{u=\hat{u}}=\B_a^i(\hat{u})=0$.\footnote{$\B_a^i(\hat{u})=0$ is stricter than $e^{\i\B_a^i(\hat{u}_a)}=1$ but we will see that the solution obtained is consistent with known results and has expected behaviour in simplifying limits of $ν$'s.} $\V$ once again turns out to be related to $F_{S^3}$ so we can easily solve it in the large $N$ limit. The second step then involves substituting this solution back in $Z_{Σ_{\fg}×S^1}$ and using the residue theorem to get the final result:
\equ{Z_{Σ_{\fg}×S^1}=∑_{\hat{u}∈\text{BAE}}\(\B(\hat{u})\)^{\fg-1} ∏_a∏_{α∈G}\(1-e^{2πα(\hat{u}_a)}\)^{1-\fg}∏_{I}∏_{ρ∈R_I}\bigg(\frac{e^{πρ(\hat{u}_I)+π\i ν_I}}{1-e^{2πρ(\hat{u}_I)+2π\i ν_I}}\bigg)^{(\fg-1)(\hat{\fn}_I-1)}\,,
\label{ZTIbaeexp}}
where the Hessian can now be rewritten as $\B=\det_{ai,bj}\frac{∂^2\V}{∂u_a^i∂u_b^j}$ and we have shifted the flavour flux with the R-charge $\hat{\fn}=(\fn+Δ)$ but we will suppress the $\hat{\hphantom{\fn}\vphantom{\fn}}$ over $\fn$ in what follows. Again, we will evaluate this final expression only in the large $N$ limit.

\subsection[\texorpdfstring{Summing Fluxes $→\V$}{Summing Fluxes → V}]{Summing Fluxes $\bm{→\V}$}\label{sec:SF}
We consider $\N=2$ quiver theories with gauge group $⊗_aU(N_a)$ now so most expressions below have a non-trivial summation $∑_{b|(a,b)∈E}$ accompanying the vector and bifundamental matter contributions when compared to similar expressions in the literature.

We begin with the BAEs which are obtained as coefficients of $\fm_a^i$ from the exponentiated form of the integrand in \eqref{ZTIgenexp}:
\eqst{0=\i\B_a^i=2πk_au_a^i+∑_j\sgn(j-i)\i π +∑_{b|(a,b)∈E}∑_j\(v'(u_a^i-u_b^j+\i ν_{(a,b)}) -v'(u_b^j-u_a^i+\i ν_{(b,a)})\) \\
+∑_{f^a}v'(u_a^i+\i ν_{f^a}) -∑_{\bar{f}^a}v'(-u_a^i+\i \bar{ν}_{f^a}).
\label{eqBAEs}}
These can be derived from the following Bethe potential $\V$ via $\B_a^i=\frac{∂\V}{∂u_a^i}$:
\eqst{\V =-\i∑_{a,i}πk_a(u_a^i)^2 +\frac{1}{2}∑_{a,i,j}π\sgn(j-i)(u_a^i-u_a^j)  -\i∑_{(a,b)∈E}∑_{i,j}\Big(v(u_a^i-u_b^j+\i ν_{(a,b)}) \\
+v(u_b^j-u_a^i+\i ν_{(b,a)})\Big) -\i∑_{a,i}∑_{f^a}v(u_a^i+\i ν_{f^a}) -\i∑_{a,i}∑_{\bar{f}^a}v(-u_a^i+\i \bar{ν}_{f^a}),
}
where we defined, in analogy to $\ell(z)$,
\equ{v(z)=\tfrac{1}{2π}\Li_2\big(e^{2\pi z}\big)+\tfrac{π}{2}z^2 -\tfrac{\pi}{12}\quad ⇒ \quad v'(z)=\Li_1(e^{2πz}) +πz\,.
\label{defvee}}
We chose $v(z)$ such that $v(0)=0$, however, $v'(z)$ is divergent at $z=0$.

To take the continuum limit, we again denote the eigenvalues of the $u_a$ matrices by $λ_{a,i}$ and assume the eigenvalue distribution for a node with $U(n_aN)$ group to be the same as before:
\equ{λ_{a,i}→λ_{a,I}(x)=N^αx+\i y_{a,I}(x) \quad \text{ (with $I=1,⋯,n_a$),}
}
with an associated eigenvalue density $ρ(x)$ normalized as $∫dx ρ(x)=1$. We convert $∑_i → N∫dxρ(x)∑_I$ and note that we again need the same number of bifundamental and anti-bifundamental matters at each edge to cancel higher order terms. To cancel potential divergent terms (as before), we are led to a constraint relating the comarks $n_a$'s and chemical potentials $ν$'s as follows (see Appendix \ref{app:DP} for details):
\eqsc{∑_a\frac{n_a^2}{2} =∑_{(a,b)∈E}(1-ν_{(a,b)}-ν_{(b,a)})n_a n_b\,.
\label{ADEconditionTTI}}
This leads to a larger class of theories than those considered in the literature whose twisted index turns out to scale as $N^{\sfrac{3}{2}}$ in the large $N$ limit. We note that for $ν_{(a,b)}+ν_{(b,a)}=\frac{1}{2}$, we get an $\wh{ADE}$ classification just like the $F_{S^3}$ as the above equation becomes equivalent to $2n_a=∑_{b|(a,b)∈E}n_b$.\footnote{As discussed in footnote \ref{ADEfootnote}, for ABJM theory and other odd $\wh{A}$ quivers, the condition can be made less strict: $ν_{(a-1,a)}+ν_{(a,a-1)}+ν_{(a,a+1)}+ν_{(a+1,a)}=1$. However, we will not discuss non-$\wh{ADE}$ constraints in detail.} This condition can also be derived from the analysis of possible superpotential terms as discussed in \cite{Benini:2015eyy,Hosseini:2016tor}. Thus, we are led to the same constraint on $α$ as before ($1+α=2-α$) implying $α=\half$ and the Bethe potential in large $N$ limit reads
\eqst{\V ≈N^{\sfrac{3}{2}}∫dx ρ(x)\Bigg[2πx ∑_{a,I}k_a y_{a,I}(x) -\frac{1}{24π^2}ρ(x)∑_{(a,b)∈E}∑_{I,J}\Big[\arg\big(\exp{2π\i(y_{a,I}-y_{b,J}+ν_{(a,b)}-\sfrac{1}{2})}\big) \\
×\Big(π^2 -\arg\big(\exp{2π\i(y_{a,I}-y_{b,J}+ν_{(a,b)}-\sfrac{1}{2})}\big)^2\Big) +(ν_{(b,a)}\text{ term})\Big] +π|x|(n_F-ν_F) \Bigg] -2π\tilde{μ}N^{\sfrac{3}{2}}\bigg(∫dx\,ρ(x) -1\bigg).
\label{BetheV}}
Here, $ν_F=∑_a∑_{\{f^a\}}n_a\(ν_{f^a}+\bar{ν}_{f^a}\)$ and we have again set $∑_an_ak_a=0$, $f^a =\bar{f}^a$. We have also added a Lagrange multiplier ($\tilde{μ}$) term to enforce the normalizability of the eigenvalue density. We can also simplify the exponent by using the constraint $ν_{(a,b)}+ν_{(b,a)}=\frac{1}{2}$, which will be employed below to derive the twisted index in terms of the Bethe potential. Notice the similarities and differences of the above expression with the expression for $F_{S^3}$ in \eqref{FS3extfin}, especially the scaling $N^{\sfrac{3}{2}}$ and missing vector contributions. This naïvely seems to suggest that $\V≈F_{S^3}$ may not hold for the larger class of theories being considered here. We will see later that it is not so. On general grounds\cite{Gulotta:2011si}, extremizing $\V$ gives (just like the free energy)
\equ{\bar{\V}=\frac{4πN^{\sfrac{3}{2}}}{3}\tilde{μ}\,.
\label{genextV}}

It turns out that the large $N$ limit of $\V$ is not enough to compute the twisted index because $\V$ has no divergences at leading order whereas the original BAEs display divergent behaviour. This behaviour follows due to bifundamental contributions involving $v'(z)$ being divergent at $z=0$ \cite{Benini:2015eyy}. Let us separate out the divergent part of \eqref{eqBAEs} but continue to denote rest of the finite terms as $\B_a^I$ and schematically introduce exponentially small corrections as follows:
\eqs{0&=\B_a^I + ∑_{b|(a,b)∈E}∑_J \Big[v'\Big(\i\(y_{a,I}(x) -y_{b,J}(x) +ν_{(a,b)}\) +e^{-N^{\sfrac{1}{2}}Y^+_{(a,I;b,J)}(x)}\Big) \nn
&\qquad\qquad -v'\Big(\i\(y_{b,J}(x) -y_{a,I}(x) +ν_{(b,a)}\) +e^{-N^{\sfrac{1}{2}}Y^-_{(a,I;b,J)}(x)}\Big) \Big] \nn
⇒\B_a^I &≈ -N^{\sfrac{1}{2}}∑_{b|(a,b)∈E}∑_J\Big[δ_{(δy_{ab,IJ}(x)+ν_{(a,b)},0)} Y^+_{(a,I;b,J)}(x) - δ_{(δy_{ab,IJ}(x)-ν_{(b,a)},0)} Y^-_{(a,I;b,J)}(x) \Big]\,,
\label{expYs}}
where $δ_{(f(x),0)}$ is the Kronecker delta symbol that equals 1 when $f(x)=0$ and 0 otherwise. We used the following large $N$ limit:
\equ{\Li_1\!\big(\text{exp}\big(2πe^{-N^{\sfrac{1}{2}}Y(x)}\big)\!\big)=-\log\!\big(1-\text{exp}\big(2πe^{-N^{\sfrac{1}{2}}Y(x)}\big)\!\big)≈-\log(-2πe^{-N^{\sfrac{1}{2}}Y(x)})≈+N^{\sfrac{1}{2}}Y(x).
}
Note that $Y^±(x)≥0$ for all $x$ so that the exponential term is subleading and is a consistency check for explicit computations. We stress that the above equation is used to extract the $Y^±(x)$ functions (while keeping track of the sign) from (naïve) equations of motion $\B_a^I$ evaluated at the saturation values of the $y(x)$'s as denoted by the $δ_{(δy(x)±ν,0)}$.

\subsection[\texorpdfstring{Integrating Holonomies $→\I$}{Integrating Holonomies → I}]{Integrating Holonomies $\bm{→\I}$}
Moving back to $Z_{Σ_{\fg}×S^1}$, we now have to derive the large $N$ limit of \eqref{ZTIbaeexp}. This limit can be taken in a similar way to the Bethe potential (see Appendix \ref{app:DP} for some details) but with fixed $α=\half$ such that the overall scaling of the index turns out to be $N^{\sfrac{3}{2}}$ as expected. To cancel the divergent terms in order to get local integrands as in the case of Bethe potential, we are led to a constraint on the flavour fluxes:
\eqsc{∑_an_a^2 =∑_{(a,b)∈E}(2-\fn_{(a,b)}-\fn_{(b,a)})n_a n_b\,.
\label{FFxcondition}}
This general constraint goes together with \eqref{ADEconditionTTI} to define a larger class of theories with $N^{\sfrac{3}{2}}$ scaling of their twisted index. Note that for $\fn_{(a,b)}+\fn_{(b,a)}=1$, we recover the $\wh{ADE}$ classification which we will impose for evaluating examples explicitly\footnote{For ABJM theory and other odd $\wh{A}$ quivers, the condition is less strict: $\fn_{(a-1,a)}+\fn_{(a,a-1)}+\fn_{(a,a+1)}+\fn_{(a+1,a)}=2$, as expected by now.}. Finally, the large $N$ limit of the twisted index reads (see appendix \ref{app:DP} for some details):
\eqs{\I &=\log|Z_{Σ_{\fg}×S^1}| ≈ (\fg-1)N^{\sfrac{3}{2}}∫dx ρ(x)\Bigg[\frac{1}{4π}ρ(x)\bigg(∑_{a,I,J}\arg\big(\exp{2π\i (y_{a,I}(x) -y_{a,J}(x) -\sfrac{1}{2})}\big)^2 \nn
&\quad -∑_{(a,b)∈E}∑_{I,J}\Big[\fn_{(b,a)}\arg\big(\exp{2π\i(y_{a,I}(x) -y_{b,J}(x)-ν_{(b,a)})}\big)^2 \Big] -(\fn_{(a,b)}\text{ term}) \bigg) \nn
&\quad +∑_{(a,b)∈E}∑_{I,J} δ_{(δy_{ab,IJ}(x)±ν_{(·,·)},0)}\fn_{(·,·)}Y^±_{(a,I;b,J)}(x) +π|x|(2n_F -\fn _F)\Bigg],
\label{indexpOS}}
where $\fn_F$ is defined similar to $ν_F$ and the conditions on $\fn_{(a,b)}$'s and $ν_{(a,b)}$'s have been used. The above expression is to be evaluated by substituting $\big\{ρ(x),y_{a,I}(x),Y^±_{(a,I;b,J)}(x)\big\}$ obtained from extremizing the Bethe potential.

\subsection[\texorpdfstring{Relating $\I→\V$}{Relating I → V}]{Relating $\bm{\I→\V}$}
We now present a simple derivation of the relation between the twisted index $\I$ and the Bethe potential $\V$. As the Bethe potential gets no contribution from vector multiplets, it seems unlikely at first that $\V$ can be directly related to $\I$. This fact is precisely what comes in handy. We augment $\V$ in \eqref{BetheV} by terms similar to the bifundamental contributions that look like adjoint contributions parameterized with $ν_a$ such that for $ν_a=0$, these adjoint terms vanish. Then, we can write off-shell:
\equ{\I = (\fg-1)∑_I\fn_I\frac{∂\V}{∂ν_I}\bigg|_{\fn_a=-1,ν_a=0},
}
where $I$ runs over all multiplets and it is understood that for vectors we set $\fn_a=-1$ and $ν_a=0$ at the end of the differentiation. This is true simply because $\V$ depends on $v(z)$ functions and $\I$ on $v'(z)$ multiplied with $(\fg-1)\fn$, though \eqref{FFxcondition} has to be used to cancel some $\frac{π}{12}$'s. The Kronecker $δ$ contributions are also included in this form, which can be shown by using the equations of motion \eqref{expYs} and chain rule for differentiation, for example,
\equ{\fn_{(a,b)}\frac{∂\V}{∂y_{a,I}}\frac{∂y_{a,I}}{∂ν_{(a,b)}}=\fn_{(a,b)}\(δ_{(δy_{ab,IJ}(x)+ν_{(a,b)},0)}Y^+_{(a,I;b,J)}(x)\)(+1)\,,}
which is what appears in \eqref{indexpOS}. The $\fn_{(b,a)}Y^-(x)$ term with proper sign also similarly follows. Thus, we have proven that the twisted index can be obtained from the Bethe potential and this relation is valid for a larger class of theories than considered in \cite{Hosseini:2016tor}, as the more general constraint \eqref{FFxcondition} was required to complete the proof.

The above formula focusses on the integrands and under certain conditions (for example, whenever definite integration and differentiation commutes), it is valid even after the integration is done (i.e., on-shell):
\equ{\bar{\I} =(\fg-1)∑_I\fn_I\frac{∂\bar{\V}}{∂ν_I}\,·
\label{simpind}}
It is understood that the index $I$ now runs only over the matter multiplets since vector $ν_a$'s are set to zero already at the level of the integrand. For (anti-)fundamental matter contributions, we can take $ν_I=n_F-ν_F$ and $\fn_I=2n_F-\fn_F$ and the above relation continues to hold. Note that we are allowed to choose a suitable basis for the $\fn$'s and $ν$'s by including even redundant combinations. Thus, to keep the expression for $\bar{\V}$ tractable, constraints on $ν_{(a,b)}$ and $\fn_{(a,b)}$ may be imposed and that makes the sum over $I$ for all bifundamentals ill-defined leading to violation of \eqref{simpind}. To understand this better, let us compare what happens to the sum $∑_{(a,b)⊕(b,a)}$ if the two constraints $ν_{(a,b)}+ν_{(b,a)}=\frac{1}{2}$ and $\fn_{(a,b)}+\fn_{(b,a)}=1$ are imposed after and before the differentiation:
\begingroup
\allowdisplaybreaks
\eqs{\text{After:}\quad ∑_I\fn_I\frac{∂\Vb}{∂ν_I}&=\fn_{(a,b)}\frac{∂\Vb(ν_{(a,b)},⋯)}{∂ν_{(a,b)}}+\fn_{(b,a)}\frac{∂\Vb(ν_{(b,a)},⋯)}{∂ν_{(b,a)}}+⋯ \nn
&=\fn_{(a,b)}\Vb'(ν_{(a,b)},⋯)+(1-\fn_{(a,b)})\Vb'(\tfrac{1}{2}-ν_{(a,b)},⋯)+⋯ \\[1cm]
\text{Before:}\quad {∑_I}'\fn_I\frac{∂\Vb}{∂ν_I}&=\fn_{(a,b)}\(\Vb'(ν_{(a,b)},⋯) -\Vb'(\tfrac{1}{2}-ν_{(a,b)},⋯)\)+⋯ \nn
&=∑_I\fn_I\frac{∂\Vb}{∂ν_I} -\Vb'(\tfrac{1}{2}-ν_{(a,b)},⋯),
\label{sumnVrel}}
\endgroup
where $∑'$ denotes sum over independent set of $ν$'s, which seems to be missing a term when compared to the full $∑$. Let us look at the following expression now:
\eqs{{∑_I}'ν_I\frac{∂\Vb}{∂ν_I} &=ν_{(a,b)}\(\Vb'(ν_{(a,b)},⋯) -\Vb'(\tfrac{1}{2}-ν_{(a,b)},⋯)\)+⋯ \nn
&=ν_{(a,b)}\Vb'(ν_{(a,b)},⋯) +(\half-ν_{(a,b)})\Vb'(\tfrac{1}{2}-ν_{(a,b)},⋯) -\half\Vb'(\tfrac{1}{2}-ν_{(a,b)},⋯)+⋯ \nn
&=∑_Iν_I\frac{∂\Vb}{∂ν_I} -\frac{1}{2}\Vb'(\tfrac{1}{2}-ν_{(a,b)},⋯),
\label{sumDVrel}}
where the last term is half of the extra term found in \eqref{sumnVrel}. Now, collecting all the terms, the general relation between twisted index and Bethe potential follows:
\eqs{\Ib =(\fg-1)∑_I\fn_I\frac{∂\Vb}{∂ν_I} &=(\fg-1)\bigg[{∑_I}'\fn_I\frac{∂\Vb}{∂ν_I} +2\bigg(∑_Iν_I\frac{∂\Vb}{∂ν_I} -{∑_I}'ν_I\frac{∂\Vb}{∂ν_I}\bigg)\bigg] \nn
&=(\fg-1)\bigg[2∑_Iν_I\frac{∂\Vb}{∂ν_I} +{∑_I}'(\fn_I-2ν_I)\frac{∂\Vb}{∂ν_I}\bigg] \nn
⇒\quad\Ib\;&=(\fg-1)\bigg[4\Vb +{∑_I}'(\fn_I-2ν_I)\frac{∂\Vb}{∂ν_I}\bigg]\,·
\label{genind}}
We used the ``homogeneous'' property of $\Vb$ such that $∑_Iν_I\frac{∂\Vb}{∂ν_I}=2\Vb$ (proven in appendix \ref{app:DP}) to write the first term. We will also see later that $4\bar{\V}[ν]=\bar{F}_{S^3}[2ν]$ for the $\wh{AD{\scriptstyle{E}}}$ quivers. The $2ν$'s here become the R-charges $Δ$'s in $F_{S^3}$ for this comparison, as can be expected from the constraints imposed on them to get $\wh{ADE}$ classification. In general, $\I$ needs to be extremized with respect to $ν$'s and critical values for $ν$'s are obtained in terms of the flavour fluxes $\fn$'s. The resulting expression $\bar{\I}(ν(\fn),\fn)$ is supposed to match the corresponding black hole entropy $S_{BH}$ as discussed in Section \ref{sec:IO}. However, for the case of universal twist, $\fn_I=2ν_I$ \cite{Bobev:2017uzs,Azzurli:2017kxo} leading to the expected simple relation for $\wh{ADE}$ quiver theories and their duals:
\equ{\text{Universal twist: }\quad S_{BH}=\bar{\I}=(\fg-1)4\bar{\V}=(\fg-1)\bar{F}_{S^3} \quad \text{ given that } \tfrac{∂\bar{\V}}{∂ν_I}≡\tfrac{∂\bar{F}_{S^3}}{∂ν_I}=0\,.
}

\vspace*{3mm}
\noindent This completes the setup for the twisted index $\I$. Let us now turn to explicit computation of the free energy of $\wh{AD}$ quivers.

\section{Free Energy and Volume}\label{sec:FEV}
In this section, we consider the $\wh{AD}$ quivers and evaluate their free energy, or equivalently the $\VY$. We will follow the algorithm developed in \cite{Crichigno:2012sk} but suitably modified for the case of general R-charges. We briefly review it here to introduce the terminology we use when writing down the explicit solutions.

\paragraph{Algorithm.} We take the principle value for the $\arg()$ functions leading to the inequalities:
\equg{0<y_{a,I}-y_{a,J}<1\,;\qquad 0<y_{a,I}-y_{b,J}+Δ_{(a,b)}<1\,,\quad -1<y_{a,I}-y_{b,J}-Δ_{(b,a)}<0\,. \\
⇒|y_{a,I}-y_{a,J}|<1\,;\qquad -Δ_{(a,b)}<y_{a,I}-y_{b,J}<Δ_{(b,a)}\,.
\label{FS3ineq}}
As discussed in previous section, we will insist $Δ_{(a,b)}+Δ_{(b,a)}=1$. Since we have pairing up of bifundamentals, while the inequalities are not violated, the contribution from these fields to \eqref{FS3extfin} simplifies:
\eqst{-∑_{\mathclap{(a,b)⊕(b,a)}}π\(2-Δ^+_{(a,b)}\)∫dx ρ(x)^2∑_{I,J}\bigg[\Big(y_{a,I}-y_{b,J}+\tfrac{Δ^-_{(a,b)}}{2}\Big)^2 +\frac{1}{12}\(3-Δ^+_{(a,b)}\)\(1-Δ^+_{(a,b)}\)\bigg] \\
=-∑_{\mathclap{(a,b)⊕(b,a)}}π∫dx ρ(x)^2∑_{I,J}\Big(y_{a,I}-y_{b,J}+\tfrac{Δ^-_{(a,b)}}{2}\Big)^2\,,
\label{bifunsimp}}
where $Δ^±_{(a,b)}=Δ_{(a,b)}±Δ_{(b,a)}$. We will also insist that all $y_{a,I}(x)-y_{a,J}(x)=0$ initially, which simplifies the vector contribution to just $∑_a∫dxρ(x)^2∑_{I,J}\frac{π}{4}$.

Extremizing $F_{S^3}$ now with respect to $y(x)$'s and $ρ(x)$, we find a solution which is consistent only in a bounded region around the origin ($x=0$). This is because as $|x|$ increases, the differences $y_{a,I}(x)-y_{b,J}(x)≡δy_{ab,IJ}(x)$ monotonically increase (or decrease), saturating at least one of the inequalities given above at some point on either side of $x=0$, which we label as $x^±_1$. This saturation is maintained beyond these points, requiring the corresponding $y_{a,I}(x)$'s to either bifurcate (for $n_a>1$) or develop a kink. Once an inequality is saturated, we have to remove one of the $y(x)$'s from the integral expression \eqref{FS3extfin} by using the saturation value and solve the revised equations of motion separately on both positive and negative side of the $x$-axis until new saturation points are encountered on both sides. This leads to pair of regions on either side of the central region (or region 1), which we will label as ``region $2^±$'' bounded by $x^±_2$ for obvious reason. This procedure needs to be iterated until either all $y(x)$'s get related or $ρ(x)=0$, determining a maximum of $∑_an_a$ regions for $\wh{A}$ quivers and $∑_an_a-1$ regions for $\wh{DE}$ quivers\footnote{We count disjointed $n^±$ regions as one single region so $\wh{A}_1$ has {\bf two} regions, even though there are four saturation points bounding {\it three} apparent regions $\{·2^-·1·2^+·\}\,$.}. Once the eigenvalue density $ρ(x)$ is determined in all the regions, the value of $μ$ is found from the normalization condition of $ρ(x)$, which gives the quantities we want via the following relations:
\equ{\bar{F}_{S^3}=\frac{4πN^{\sfrac{3}{2}}}{3}μ\,; \qquad\qquad \frac{\VY}{\Vol(S^7)}=\frac{1}{8μ^2}\,·
}
We combined the former equation with \eqref{FS3vol} to get the latter.

\subsection[\texorpdfstring{$\wh{A}_m$ Revisited}{Am Revisited}]{$\bm{\wh{A}_m}$ Revisited}
We review the $\wh{A}$ quivers dealt succinctly in \cite{Gulotta:2011aa}. The above discussion applies to this case just by setting the values of $I,J=1$. The contribution from bifundamentals \eqref{bifunsimp} can be rewritten as $\tilde{F}$ given by eq. (4.2) of \cite{Gulotta:2011aa}. The solution for free energy is given in terms of the area of the following polygon:
\equ{\P=\Big\{(s,t)∈\bR^2\Big| {\textstyle ∑_{a=1}^{m+1}}|t +q_a s| +c_1 t+ c_2 s ≤1\Big\}; \quad c_1≡∑_{\mathclap{(a,b)∈E}}Δ_{(a,b)}^-,\quad c_2≡∑_{\mathclap{(a,b)∈E}}q_a Δ_{(a,b)}^-.
\label{coneA}}
The redefined CS levels $q_a$ are constrained parameters obeying $∑_{a=1}^{m+1}q_a=0$ and are related to $k_a$'s as follows:
\equ{q_a = k_a-k_{a+1}\,,\; a=1,⋯,m\,;\quad q_{m+1}=k_{m+1} -k_1\,.
\label{k2qA}}
The $\Ar(\P)$ is related to $∫dx ρ(x)$ such that we get\footnote{It is a fun exercise to show that the definition of $\P$ as given in \eqref{coneA} can be ``integrated'' to get precisely the area of $\P$ as given in \eqref{genvolA} \cite{Gulotta:2011si}.}
\equ{\frac{\VY}{\Vol(S^7)}=\frac{1}{8μ^2}=\frac{1}{2}\Ar(\P)=\frac{1}{4}∑_{a=1}^{m+1}\left[\frac{|γ_{a,a+1}|}{\s_a\s_{a+1}}+\frac{|γ_{a,a+1}|}{\s_{a+m+1}\s_{a+m+2}}\right].
\label{genvolA}}
This reduces to the correct $\N=3$ expression when all $Δ_{(a,b)}=\half$ as can be directly checked from the definition of $\s$'s:
\eqsc{β_a=\spmat{1 \\q_a}\,\text{ for } a=1,⋯,m+1\,,\; β_{m+2}=-β_1\;; \qquad γ_{a,b}=β_a\wedge β_b\,; \nn
\s_a=∑_{b=1}^{m+1}\left[|γ_{a,b}|+γ_{b,a}Δ_{(b,b+1)}^-\right]\,,\quad \s_{a+m+1}=∑_{b=1}^{m+1}\left[|γ_{a,b}| -γ_{b,a}Δ_{(b,b+1)}^-\right].
\label{bgsdefA}}

A trivial example to check the above formulas is $\wh{A}_1$ quiver (consider the ordering $q_1≥0≥q_2$ and $q_1=\frac{k}{2}$):
\equ{\frac{\VY}{\Vol(S^7)} =\frac{1}{4}\pmat{\frac{2q_1}{(4q_1Δ^A_{(2,1)})(4q_1Δ^B_{(1,2)})} +\frac{2q_1}{(4q_1Δ^B_{(1,2)})(4q_1Δ^B_{(2,1)})} \\ +\frac{2q_1}{(4q_1Δ^B_{(2,1)})(4q_1Δ^A_{(1,2)})} +\frac{2q_1}{(4q_1Δ^A_{(1,2)})(4q_1Δ^A_{(2,1)})}} =\frac{1}{32Δ^A_{(1,2)}Δ^B_{(1,2)}Δ^A_{(2,1)}Δ^B_{(2,1)}\,q_1}\,·
}
This expression appears in literature a lot and it can be straightforwardly checked that it reproduces the correct $\frac{1}{k}$ for ABJM theory when all $Δ$'s equal $\half$. A slightly non-trivial example is $\wh{A}_3$ but we will discuss it for twisted index in the next section.\\

\noindent Let us move on to the $\wh{D}$ quivers now (specifically $\wh{D}_4$ which is related to $\wh{A}_3$ via unfolding procedure in the $\N=3$ case\cite{Gulotta:2012yd,Crichigno:2012sk}).

\subsection[\texorpdfstring{$\wh{D}_4$ Solved}{D₄ Solved}]{$\bm{\wh{D}_4}$ Solved}\label{D4freeEn}
We give the detailed solution for the $\wh{D}_4$ quiver here and to make the expressions easier to read, we do a bit of housekeeping first. Let us redefine the five constrained CS levels $k$'s to four unconstrained variables $p$'s as follows:
\equ{k_1=-(p_1+p_2),\quad k_2=p_1-p_2,\quad k_3=p_3-p_4,\quad k_4=p_3+p_4,\quad k_5=p_2-p_3\,.
}
We will also suppress the second index on the four $y_{a,1}$ with $a=1,⋯,4$. Furthermore, we introduce a ``vector'' of R-charges:
\equ{α_b(Δ^-)=\Big\{\tfrac{1}{2}(\Delta^-_{(1,5)}-\Delta^-_{(2,5)}), \tfrac{1}{2}(\Delta^-_{(1,5)}+\Delta^-_{(2,5)}), -\tfrac{1}{2}(\Delta^-_{(4,5)}+\Delta^-_{(3,5)}),-\tfrac{1}{2}(\Delta^-_{(4,5)}-\Delta^-_{(3,5)})\Big\},
}
which will appear in a combination $∑_{b=1}^4p_bα_b(Δ^-)≡p·α(Δ^-)$ below. For generic $p$'s, there are going to be 5 regions consisting of one central region spanning both negative and positive side of the $x$-axis and 4 pairs of disjointed regions beyond the central one as explained in the algorithm above. Let us now enumerate the solution in each region for a particular ordering $p_1≥p_2≥p_3≥p_4≥0$. 

\paragraph{Region 1:} $-\tfrac{2 \mu}{4 (p_1 + p_2) - 2 p·α(Δ^-)}\leq x \leq \tfrac{2 \mu}{4 (p_1 + p_2) + 2 p·α(Δ^-)}$
\eqscn{\rho(x) = \tfrac{1}{2}\mu - \tfrac{1}{2}xp·α(Δ^-)\,; \\
y_{1}-y_{5,2} = -\tfrac{1}{2}\Delta^{-}_{(1,5)} + \tfrac{ x (p_1+p_2) }{- \mu +x p·α(Δ^-)}\,, \qquad
y_{2}-y_{5,2} = -\tfrac{1}{2}\Delta^{-}_{(2,5)} + \tfrac{ x (-p_1+p_2) }{- \mu +x p·α(Δ^-)}\,, \\
y_{3}-y_{5,2} = -\tfrac{1}{2}\Delta^{-}_{(3,5)} + \tfrac{ x (-p_3+p_4) }{- \mu +x p·α(Δ^-)}\,, \qquad
y_{4}-y_{5,2} = -\tfrac{1}{2}\Delta^{-}_{(4,5)} - \tfrac{ x (p_3+p_4) }{- \mu +x p·α(Δ^-)}\,, \qquad y_{5,1}-y_{5,2} =0\,.
}

\paragraph{Region $\bm{2^-}$:} $-\tfrac{2\mu}{4 p_1 - 2 p·α(Δ^-)}\leq x \leq - \tfrac{2 \mu}{4 (p_1 + p_2) - 2 p·α(Δ^-)}$
\eqscn{\rho (x) = \tfrac{1}{2}\mu - \tfrac{1}{2}xp·α(Δ^-) \,; \\
y_{1}-y_{5,2} = 1 -\Delta_{(1,5)}\,, \qquad 
y_{2}-y_{5,2} = \tfrac{1}{2}\big(1 -\Delta^-_{(2,5)}\big) - \tfrac{2 x p_1}{- \mu + x p·α(Δ^-)}\,, \\
y_{3}-y_{5,2} = \tfrac{1}{2}\big(1 -\Delta^-_{(3,5)}\big) - \tfrac{x(p_1 +p_2 +p_3 -p_4) }{-\mu +x p·α(Δ^-)}\,, \qquad
y_{4}-y_{5,2} = \tfrac{1}{2}\big(1 -\Delta^-_{(4,5)}\big) - \tfrac{x(p_1 +p_2 +p_3 +p_4) }{-\mu +x p·α(Δ^-)}\,, \\
y_{5,1}-y_{5,2} = 1 -\tfrac{2 x(p_1 + p_2) }{- \mu + x p·α(Δ^-)}\,·
}
\paragraph{Region $\bm{2^+}$:} $\tfrac{2 \mu}{4 (p_1 + p_2) + 2 p·α(Δ^-) }\leq x \leq \tfrac{2 \mu}{4 p_1  + 2 p·α(Δ^-)}$
\eqscn{\rho (x) = \tfrac{1}{2}\mu - \tfrac{1}{2}xp·α(Δ^-)\,; \\
y_{1}-y_{5,2} = -\Delta_{(1,5)}\,, \qquad
y_{2}-y_{5,2} = -\tfrac{1}{2}\big(1 +\Delta^-_{(2,5)}\big) -\tfrac{2 x p_1}{- \mu + xp·α(Δ^-)}\,, \\
y_{3}-y_{5,2} = \tfrac{1}{2}\big(1 -\Delta^-_{(3,5)}\big) -\tfrac{x(p_1 + p_2 + p_3- p_4) }{-\mu + xp·α(Δ^-)}\,, \qquad
y_{4}-y_{5,2} = \tfrac{1}{2}\big(1 -\Delta^-_{(4,5)}\big) -\tfrac{x(p_1 + p_2 + p_3+ p_4) }{-\mu + xp·α(Δ^-)}\,, \\
y_{5,1}-y_{5,2} = 1 -\tfrac{2 x(p_1 + p_2) }{- \mu + xp·α(Δ^-)}\,·
}

\paragraph{Region $\bm{3^-}$:} $-\tfrac{2\mu}{2(p_1 + p_2 + p_3 + p_4) - 2 p·α(Δ^-)}\leq x \leq  - \tfrac{2\mu}{4 p_1 - 2 p·α(Δ^-)}$
\eqscn{\rho (x) = \mu +xp_1 - x p·α(Δ^-)\,; \\
y_{1}-y_{5,2} = 1 -\Delta_{(1,5)}\,, \qquad 
y_{2}-y_{5,2} = -\Delta_{(2,5)}\,, \qquad
y_{3}-y_{5,2} = -\tfrac{1}{2} \Delta^-_{(3,5)} -\tfrac{x\left(p_2 +p_3-p_4 \right) }{-2 \mu -2x p_1 + 2 x p·α(Δ^-)}\,, \\
y_{4}-y_{5,2} = -\tfrac{1}{2} \Delta^-_{(4,5)} -\tfrac{x\left(p_2 +p_3+p_4 \right) }{-2 \mu -2x p_1 + 2 x p·α(Δ^-)}\,, \qquad
y_{5,1}-y_{5,2} = -\tfrac{2x p_2  }{-2 \mu -2x p_1 + 2 x p·α(Δ^-)}\,·
}
\paragraph{Region $\bm{3^+}$:} $\tfrac{2 \mu}{4 p_1 + 2 p·α(Δ^-)}\leq x \leq \tfrac{2 \mu}{2(p_1+p_2+p_3+p_4) + 2p·α(Δ^-)}$
\eqscn{\rho (x) = \mu -xp_1 -x p·α(Δ^-)\,; \\
y_{1}-y_{5,2} = -\Delta_{(1,5)}\,, \qquad
y_{2}-y_{5,2} = 1 -\Delta_{(2,5)}\,, \qquad
y_{3}-y_{5,2} = -\tfrac{1}{2} \Delta^-_{(3,5)} -\tfrac{x\left(p_2 +p_3-p_4 \right)}{-2 \mu + 2x p_1 + 2 x p·α(Δ^-)}\,, \\
y_{4}-y_{5,2} = -\tfrac{1}{2} \Delta^-_{(4,5)} -\tfrac{x\left(p_2 +p_3+p_4 \right)}{-2 \mu + 2x p_1 + 2 x p·α(Δ^-)}\,, \qquad
y_{5,1}-y_{5,2} = -\tfrac{2x p_2}{-2 \mu + 2x p_1 + 2 x p·α(Δ^-)}\,·
}

\paragraph{Region $\bm{4^-}$:} $-\tfrac{2\mu}{2(p_1+p_2+p_3-p_4)-2p·α(Δ^-)} \leq x \leq - \tfrac{2\mu}{2(p_1+p_2+p_3+p_4)-2p·α(Δ^-)}$
\eqscn{\rho (x) = \tfrac{3}{2} \mu+\tfrac{1}{2}x (3 p_1 + p_2 + p_3 + p_4) - \tfrac{3}{2} xp·α(Δ^-)\,; \\
y_{1}-y_{5,2} = 1 -\Delta_{(1,5)}\,, \qquad 
y_{2}-y_{5,2} = -\Delta_{(2,5)}\,, \\
y_{3}-y_{5,2} = -\tfrac{1}{6} -\tfrac{1}{2} \Delta^-_{(3,5)} +\tfrac{ 2x (2p_4-p_2- p_3) }{-9 \mu -3x(3 p_1 + p_2 + p_3 + p_4) + 9 x p·α(Δ^-)}\,, \qquad
y_{4}-y_{5,2} = -\Delta_{(4,5)}\,, \\
y_{5,1}-y_{5,2} = -\tfrac{1}{3} +\tfrac{2x(-2 p_2+p_3+p_4)}{-9 \mu - 3x (3 p_1 + p_2 + p_3 + p_4)+9 x p·α(Δ^-)}\,·
}
\paragraph{Region $\bm{4^+}$:} $\tfrac{2 \mu}{2(p_1+p_2+p_3+p_4)+2p·α(Δ^-)} \leq x \leq \tfrac{2 \mu}{2(p_1+p_2+p_3-p_4)+2p·α(Δ^-)}$
\eqscn{\rho (x) = \tfrac{3}{2} \mu - \tfrac{1}{2}x(3 p_1 + p_2 + p_3 + p_4) - \tfrac{3}{2}x p·α(Δ^-)\,; \\
y_{1}-y_{5,2} = -\Delta_{(1,5)}\,, \qquad
y_{2}-y_{5,2} = 1 -\Delta_{(2,5)}\,, \\
y_{3}-y_{5,2} = \tfrac{1}{6} -\tfrac{1}{2} \Delta^-_{(3,5)} +\tfrac{2 x (2p_4-p_2-p_3) }{-9 \mu + 3x(3 p_1 + p_2 + p_3 + p_4)+ 9x p·α(Δ^-)}\,, \qquad
y_{4}-y_{5,2} = 1 -\Delta_{(4,5)}\,, \\
y_{5,1}-y_{5,2} = \tfrac{1}{3}+\tfrac{2x(-2 p_2+p_3+p_4)}{-9 \mu + 3x(3p_1 +p_2 + p_3 +p_4) + 9 x p·α(Δ^-)}\,·
}

\paragraph{Region $\bm{5^-}$:} $-\tfrac{2\mu}{2(p_1 + p_2)-2p·α(Δ^-)} \leq x \leq -\tfrac{2\mu}{2(p_1+p_2+p_3-p_4)-2p·α(Δ^-)}$
\eqscn{\rho (x) = 2\mu +x(2 p_1 + p_2 + p_3) -2 x p·α(Δ^-)\,; \\
y_{1}-y_{5,2} = 1 - \Delta_{(1,5)}\,, \qquad 
y_{2}-y_{5,2} = -\Delta_{(2,5)}\,, \qquad
y_{3}-y_{5,2} = -\Delta_{(3,5)}\,, \\
y_{4}-y_{5,2} = -\Delta_{(4,5)}\,, \qquad
y_{5,1}-y_{5,2} = -\tfrac{1}{2} -\tfrac{x (p_2 - p_3)}{-4 \mu -2x(2 p_1 + p_2 + p_3) +4xp·α(Δ^-)}\,·
}
Finally, the last saturation occurs at the end of this region with $y_{5,1}-y_{5,2}=-1$.
\paragraph{Region $\bm{5^+}$:} $\tfrac{2\mu}{2(p_1+p_2+p_3-p_4)+2p·α(Δ^-)} \leq x \leq \tfrac{2 \mu}{2(p_1+p_2)+2p·α(Δ^-)}$
\eqscn{\rho (x) = 2\mu -x(2p_1 + p_2 + p_3) -2 xp·α(Δ^-)\,; \\
y_{1}-y_{5,2} = -\Delta_{(1,5)}\,, \qquad
y_{2}-y_{5,2} = 1 -\Delta_{(2,5)}\,, \qquad
y_{3}-y_{5,2} = 1 -\Delta_{(3,5)}\,, \\
y_{4}-y_{5,2} = 1 -\Delta_{(4,5)}\,, \qquad
y_{5,1}-y_{5,2} = \tfrac{1}{2} -\tfrac{x (p_2 - p_3)}{-4 \mu +2x (2 p_1 +p_2 + p_3) +4 x p·α(Δ^-)}\,·
}
Finally, the last saturation occurs at the end of this region with $y_{5,1}-y_{5,2}=1$.\\

\fig{h!}{\includegraphics[scale=1]{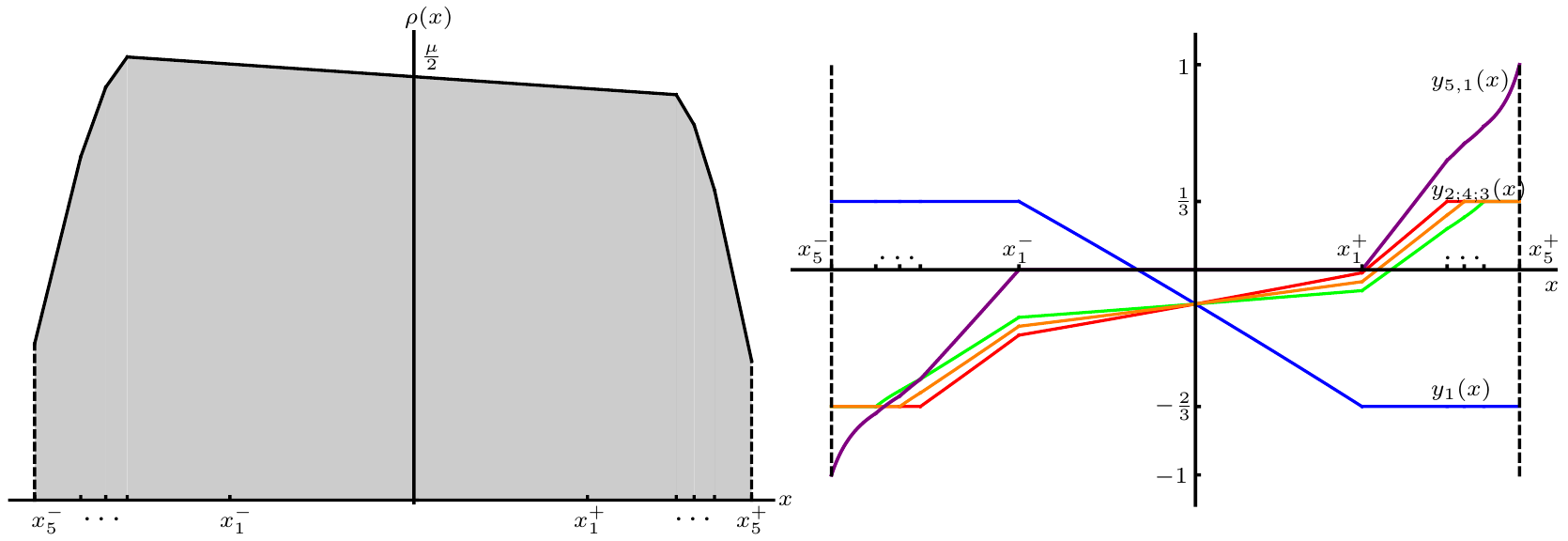}
\caption{Eigenvalue density $ρ(x)$ and distributions $y_{a,I}(x)$ for $\wh{D}_4$ quiver ($y_{5,2}(x)=0$).}
\label{figD4rhoys}}
To get a feel for these expressions for $ρ(x)$ and $y(x)$'s, we have plotted them in Figure \ref{figD4rhoys} using the numerical values: $p_1=15$, $p_2=8$, $p_3=4$, $p_4=1$ and all $Δ$'s equal to $\frac{2}{3}$. With the $ρ(x)$ known in all the regions, we can just use the normalization condition $∫dx ρ(x)=1$ to get $\frac{1}{μ^2}$, which is directly related to the $\VY$. As with the $\wh{A}$ quiver, this volume can be recast as a polygon's area and for $\wh{D}_4$ quiver, this polygon turns out to be 
\equg{\P=\Big\{(s,t)∈\bR^2\Big|{\textstyle ∑_{a=1}^4}\(|t +p_a s| +|t -p_a s|\) -4|t| +2\,p·α(Δ^-)s ≤1\Big\}\,, \\
\text{with }\; \frac{\VY}{\Vol(S^7)}=\frac{1}{4}\Ar(\P)\,.
\label{coneD4}}
For the above mentioned numerical values, the polygon is shown in Figure \ref{figD4poly} with $\frac{1}{4}\Ar(\P)=\frac{1992856091659101}{388764834312025600}≈0.005$. This value matches $\frac{\VY}{\Vol(S^7)}=\frac{1}{8μ^2}$ exactly. Also, note that this construction is valid for all possible orderings and signs of $p$'s.
\fig{h!}{\includegraphics[width=0.99\textwidth]{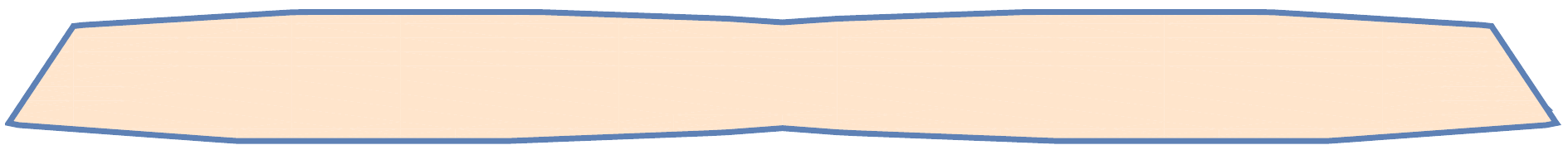}
\caption{Polygon $\P$ for $\wh{D}_4$ quiver. ($s-t$ coordinate system rotated by $\frac{π}{2}$.)}
\label{figD4poly}}

We can, of course, write the explicit volume for $\wh{D}_4$ here but instead we prefer to give the explicit expression for general $\wh{D}_n$ quiver directly.

\subsection[\texorpdfstring{$\wh{D}_n$ Result}{Dn Result}]{$\bm{\wh{D}_n}$ Result}
Given the result for $\wh{D}_4$ quiver above and the known result for $\N=3$ $\wh{D}_n$ quivers\cite{Crichigno:2012sk}, we conjecture the polygon for $\N=2$ $\wh{D}_n$ quivers to be:
\eqsc{\P=\Big\{(s,t)∈\bR^2\Big|∑_{a=1}^n\(|t +p_a s| +|t -p_a s|\) -4|t| +c\,s ≤1\Big\}\,;\quad c≡∑_{b=1}^n(2p_b)α_b(Δ^-)\,, \label{coneD} \\
α_b(Δ^-) =\Big\{\half\big(Δ^-_{(1,5)} -Δ^-_{(2,5)}\big), \half\big(Δ^-_{(1,5)}+Δ^-_{(2,5)}\big), Δ^-_{(5,6)}, ⋯, Δ^-_{(n,n+1)}, \nn
\qquad -\half\big(Δ^-_{(4,n+1)}+Δ^-_{(3,n+1)}\big), -\half\big(Δ^-_{(4,n+1)} -Δ^-_{(3,n+1)}\big)\Big\}.
\label{defalpha}
}
For generic $n$, the $p$'s are related to the CS levels as follows:
\eqsc{k_1=-(p_1+p_2),\quad k_2=p_1-p_2,\quad k_3=p_{n-1} -p_n,\quad k_4=p_{n-1}+p_n, \nn
k_i=p_{i-3}-p_{i-2};\quad i=5,⋯,n+1\,.
}

Note the difference with \eqref{coneA} for $\wh{A}$ quivers which requires two constants. As explained in \cite{Gulotta:2011aa}, this is due to the two $U(1)$ isometries of the toric $\wh{A}$ quivers so it makes sense that for the case of non-toric $\wh{D}$ quivers which has only one $U(1)$ isometry, we see only one constant in the polygon formula \eqref{coneD}.

One can verify that this polygon's area gives the general volume formula corresponding to $\wh{D}_n$ quivers:
\equ{\frac{\VY}{\Vol(S^7)}=\frac{1}{8μ^2}=\frac{1}{4}\Ar(\P)=\frac{1}{4}∑_{a=0}^n \left[\frac{|γ_{a,a+1}|}{\bar{\s}^+_a\bar{\s}^+_{a+1}}+\frac{|γ_{a,a+1}|}{\bar{\s}^-_{a}\bar{\s}^-_{a+1}}\right],
\label{genvolD}}
which we have explicitly checked for $\wh{D}_5,⋯,\wh{D}_{10}$.\footnote{It is interesting to note that the $±$ structure in \eqref{genvolD} produces independent terms, which is in contrast to the expression \eqref{genvolA} of $\wh{A}$ quivers, where the analogous $\s^+$ and $\s^-$ terms produce one mixed term. However, that is just an artifact of the way we have defined $β$'s. $β_0$ is quite redundant if we realize $\frac{1}{\bar{\s}^+_0\bar{\s}^+_1}+\frac{1}{\bar{\s}^-_0\bar{\s}^-_1}=\frac{|γ_{-1,1}|}{\bar{\s}^-_1\bar{\s}^+_1}·$} The definitions of various quantities are slightly elaborate here:
\eqsc{β_0=\spmat{0 \\ 1},\; β_{±a}=\spmat{1 \\ ±p_a}\,\text{ for } a=1,⋯,n,\; β_{n+1}=\spmat{1 \\ 0}; \qquad γ_{a,b}=β_a\wedge β_b\,; \nn
\bar{\s}^±_a=∑_{b=1}^n\left[|γ_{a,b}| +|γ_{a,-b}| ±(γ_{a,b} -γ_{a,-b})α_b(Δ^-)\right] -4|γ_{a,n+1}|\,.
\label{bgsdefD}}
The combination $(γ_{a,b} -γ_{a,-b})=2p_b$ for $a≠0$ is used to show similarity with the definitions for $\wh{A}$ quivers in \eqref{bgsdefA}, otherwise it is a simple factor defining $c$ in \eqref{coneD}.\\

\noindent This completes the free energy or dual volume computation of $\wh{AD}$ quivers. Let us now continue with the computation of their twisted indices.

\section{Twisted Index and Entropy}\label{sec:TIE}
We will again work on $\wh{AD}$ quivers and first evaluate the Bethe potential \eqref{BetheV} and then the index \eqref{indexpOS} (equivalently, dual black hole entropy). We will follow the same algorithm used to evaluate $F_{S^3}$ but start with the reduced set of inequalities as VM do not contribute to $\V$:
\equ{0<y_{a,I}-y_{b,J}+ν_{(a,b)}<1\,; \qquad -1<y_{a,I}-y_{b,J}-ν_{(b,a)}<0\,.
}
As discussed before, we will insist $ν_{(a,b)}+ν_{(b,a)}=\half$. Since we have pairing up of bifundamentals, while the inequalities are not violated, the contribution from these fields simplify to
\eqst{-∑_{\mathclap{(a,b)⊕(b,a)}}π\(1-ν^+_{(a,b)}\)∫dx ρ(x)^2∑_{I,J}\bigg[\Big(y_{a,I}-y_{b,J}+\tfrac{ν^-_{(a,b)}}{2}\Big)^2 -\frac{1}{12}ν^+_{(a,b)}\(2-ν^+_{(a,b)}\)\bigg] \\
=-∑_{\mathclap{(a,b)⊕(b,a)}}\frac{π}{2}∫dx ρ(x)^2∑_{I,J}\bigg[\Big(y_{a,I}-y_{b,J}+\tfrac{ν^-_{(a,b)}}{2}\Big)^2 -\frac{1}{16}\bigg]\,,
\label{Vbifunsimp}}
where $ν^±_{(a,b)}=ν_{(a,b)}±ν_{(b,a)}$. Comparing \eqref{FS3extfin} and \eqref{bifunsimp} in the central region (i.e., $y_{a,I}=y_{a,J}$) with \eqref{BetheV} and \eqref{Vbifunsimp}, we find that the two expressions (whether off-shell or on-shell) are same up to the scalings given in Table \ref{tabScal}.
\tabl{h!}{
\begin{tabular}{|c@{$\qquad→\qquad$}c|}
\hline
Free Energy & Bethe Potential \\
\hline\hline
$F_{S^3}$ & $4\V$ \\
$μ$ & $4\tilde{μ}$ \\
\hline
$Δ$ & $2ν$ \\
$y$ & $2y$ \\
$x$ & $2x$ \\
$ρ$ & $\frac{1}{2}ρ$ \\
\hline
\end{tabular}
\caption{Scaling different parameters to relate $F_{S^3}$ and $\V$. Note that these scalings are different from those of \cite{Hosseini:2016tor}.}
\label{tabScal}
}

Once an inequality is saturated, we have to use the general expression involving $\arg()$ functions. This step is to be taken much more seriously here than the case of free energy because moving away from the central region generates terms like $(y_{a,I}-y_{a,J})^2$ leading to new inequalities:
\equ{-\half <y_{a,I}-y_{a,J}<0 \quad \text{ or } \quad 0<y_{a,I}-y_{a,J}<\half\,,
}
which can drastically affect the evaluation of $\V$ in the new regions. This process of generation of new terms and inequalities that look like coming from vector contributions of $F_{S^3}$ means that $\V$ can indeed be related to $F_{S^3}$ in all the regions, not just in the central region, even though these two expressions seemed very different for $\wh{DE}$ quivers in subsection \ref{sec:SF}. In fact, using the scalings given in Table \ref{tabScal}, we can verify that it is indeed so allowing us to use the results for $F_{S^3}$ to write down $\V$ for the same quiver.

As far as saturation points, $ρ(x)$ and $y_{a,I}(x)$ are concerned, we can get them from the similar computations already done for $F_{S^3}$ but to get the divergent contributions $Y^±(x)$, we need to perform one more step during extremization of $\V$ in different regions. This step is to substitute the solutions of each region $n^±$ in the equations of motion $\B_a^I$ found in the region 1. Of course, $\B_a^I≠0$ in other regions but provide the divergent contributions $Y^±(x)$'s via \eqref{expYs}. One technicality is that the $\B_a^I$ of \eqref{expYs} are related to the equations of motion obtained from $\V$ via $\frac{∂\V}{∂y_{a,I}}=Nρ(x)\B_a^I$. This step needs a slight modification as discussed in subsection \ref{sec:D4s}.

\subsection[\texorpdfstring{$\wh{A}_3$ Solved}{A₃ Solved}]{$\bm{\wh{A}_3}$ Solved}\label{sec:A3s}
As far as we know, only theories like $\wh{A}_1$ quiver whose matrix models involve just 2 regions have been discussed in the literature. So we improve the situation by considering a non-trivial example explicitly for $\wh{A}$ quivers: $\wh{A}_3$, whose matrix model involves 4 regions. Let us set up some notation before presenting the explicit solution. We use the redefined CS variables following \eqref{k2qA} with the given ordering: $q_1>q_2>0>q_3$ and $q_4=-∑_{a=1}^3q_a$. We will again suppress the second index on the four $y_{a,1}$ with $a=1,⋯,4$ and introduce two short-hand notations:
\equa{\Sigma_{\nu} &= \nu^-_{(1,2)} +\nu^-_{(2,3)} +\nu^-_{(3,4)} +\nu^-_{(4,1)}\,, \\
\alpha_b(\nu^-) &= \Big\{\big(\nu^-_{(1,2)}-\nu^-_{(4,1)}\big), \big(\nu^-_{(2,3)}-\nu^-_{(4,1)}\big), \big(\nu^-_{(3,4)}-\nu^-_{(4,1)}\big)\Big\}\,,
}
which will appear in a combination $∑_{b=1}^3q_bα_b(ν^-)≡q·α(ν^-)$ below. 

\paragraph{Region 1:} $-\tfrac{2 \tilde{\mu}}{2 q_1 +q_1\Sigma_{\nu} -  q \cdot\alpha(\nu^-)}\leq x \leq \tfrac{2 \tilde{\mu}}{2 q_1 -q_1\Sigma_{\nu}+q\cdot\alpha(\nu^-)}$
\eqscn{\rho(x) = - \tfrac{32 \tilde{\mu} - 16 x q\cdot\alpha(\nu^-)}{(\Sigma_{\nu}-2)(\Sigma_{\nu}+2 )}\,; \\
y_a - y_{a+1} = \tfrac{2\tilde{\mu} (\Sigma_{\nu}-4\nu ^-_{(a,a+1)})+x[q_a(\Sigma_{\nu}-2)(\Sigma_{\nu}+2)-q\cdot\alpha(\nu^-)(\Sigma_{\nu}-4\nu ^-_{(a,a+1)})]}{16 \tilde{\mu} - 8x q\cdot\alpha(\nu^-)} \,,\quad a=1,2,3\,.
}

\paragraph{Region $\bm{2^-}$:} $-\tfrac{2\tilde{\mu}}{2 (q_1+q_2+q_3)-(q_1+q_2+q_3)\Sigma_{\nu}  -  q\cdot\alpha(\nu^-)} \leq x \leq - \tfrac{2 \tilde{\mu}}{2 q_1 +q_1\Sigma_{\nu} - q\cdot\alpha(\nu^-)}$
\eqscn{\rho (x) = - \tfrac{24\tilde{\mu} - 4x [q_1(\Sigma_{\nu}-2)+3 q\cdot\alpha(\nu^-)]  }{(\Sigma_{\nu}-2)(\Sigma_{\nu}+1)} \,; \qquad y_1-y_2 = \tfrac{1}{2}-\nu_{(1,2)}\,, \\
y_a-y_{a+1} = \tfrac{2\tilde{\mu} (2\Sigma_{\nu}-6\nu^-_{(a,a+1)}-1)+x[(q_1 (1+2\nu^-_{(a,a+1)}) + 2q_a (\Sigma_{\nu}+1))(\Sigma_{\nu}-2) -q\cdot\alpha(\nu^-)(2\Sigma_{\nu}-6\nu^-_{(a,a+1)}-1)]}{24 \tilde{\mu}-4 x q_1 (\Sigma_{\nu}-2)-12 xq\cdot\alpha(\nu^-) }\,,\quad a=2,3\,; \\
Y^-_{(1;2)} = - \tfrac{4 \pi \tilde{\mu} +2 \pi x[q_1(\Sigma_{\nu}+2)-q\cdot\alpha(\nu^-)]}{\Sigma_{\nu}+1}\,\cdot 
}
\paragraph{Region $\bm{2^+}$:} $\tfrac{2 \tilde{\mu}}{2 q_1 -q_1\Sigma_{\nu}+ q\cdot\alpha(\nu^-) }\leq x \leq \tfrac{2 \tilde{\mu}}{2 (q_1+q_2+q_3)+(q_1+q_2+q_3)\Sigma_{\nu} +  q\cdot\alpha(\nu^-) }$
\eqscn{\rho(x) =- \tfrac{24\tilde{\mu} - 4x [q_1(\Sigma_{\nu}+2)+3 q\cdot\alpha(\nu^-)]  }{(\Sigma_{\nu}+2)(\Sigma_{\nu}-1)}\,; \qquad y_1-y_2 = -\nu_{(1,2)}\,, \\
y_a-y_{a+1} = \tfrac{2\tilde{\mu} (2\Sigma_{\nu}-6\nu^-_{(a,a+1)}+1)+x[(q_1(-1+2\nu^-_{(2,3)})+2q_a(\Sigma_{\nu}-1))(\Sigma_{\nu}+2) -q\cdot\alpha(\nu^-)(2\Sigma_{\nu}-6\nu^-_{(a,a+1)}+1)]}{24 \tilde{\mu}-4 x q_1 (\Sigma_{\nu}+2)-12 xq\cdot\alpha(\nu^-)}\,,\quad a=2,3\,; \\
Y^+_{(1;2)} = \tfrac{ 4 \pi \tilde{\mu} +2 \pi x[q_1(\Sigma_{\nu}-2)-q\cdot\alpha(\nu^-)]}{\Sigma_{\nu}-1}\,\cdot 
}

\paragraph{Region $\bm{3^-}$:} $-\tfrac{2\tilde{\mu}}{q_1+q_2  +q_3\Sigma_{\nu}-q\cdot\alpha(\nu^-)}\leq x \leq  -\tfrac{2\tilde{\mu}}{2 (q_1+q_2+q_3) -(q_1+q_2+q_3)\Sigma_{\nu}-  q\cdot\alpha(\nu^-) }$
\eqscn{\rho (x) = \tfrac{-16\tilde{\mu} - 4x [2 q_1+(q_2+q_3)(1+\Sigma_{\nu})-2 q\cdot\alpha(\nu^-)]}{(\Sigma_{\nu}+1)(\Sigma_{\nu}-1)}\,; \qquad y_1-y_2 = \tfrac{1}{2}-\nu_{(1,2)}\,,\qquad y_4-y_1 = -\nu_{(4,1)}\,, \\
y_3-y_4 = \tfrac{4\tilde{\mu}(\Sigma_{\nu}-2 \nu^-_{(3,4)})+x[2q_1 (\Sigma_{\nu}-2 \nu^-_{(3,4)})+(q_2(1-2\nu^-_{(3,4)}) -q_3(1-2\Sigma_{\nu} +2\nu^-_{(3,4)}))(\Sigma_{\nu}+1) -2q \cdot \alpha(\nu^-)(\Sigma_{\nu}-2 \nu^-_{(3,4)})]}{16\tilde{\mu} + 8x q_1+4x(q_2+q_3)(\Sigma_{\nu}+1)-8xq \cdot \alpha(\nu^-)}\,; \\
Y^-_{(1;2)} = \tfrac{-4 \pi \tilde{\mu} -2 \pi x[q_1(\Sigma_{\nu}+2)-q\cdot\alpha(\nu^-)]}{\Sigma_{\nu}+1}\,,\qquad Y^+_{(4;1)} = \tfrac{ 4 \pi \tilde{\mu} - 2 \pi x[(q_1+q_2+q_3)(\Sigma_{\nu}-2)+q\cdot\alpha(\nu^-)]}{\Sigma_{\nu}-1}\,\cdot 
}
\paragraph{Region $\bm{3^+}$:} $\tfrac{2\tilde{\mu}}{2(q_1+q_2+q_3)  +(q_1+q_2+q_3)\Sigma_{\nu}+q\cdot\alpha(\nu^-)}\leq x \leq  \tfrac{2\tilde{\mu}}{ q_1+q_2 -q_2\Sigma_{\nu}+  q\cdot\alpha(\nu^-)}$
\eqscn{\rho (x) = \tfrac{-16\tilde{\mu} +4 x [2 q_1-(q_2+q_3)(\Sigma_{\nu}-1)+2 q\cdot\alpha(\nu^-)]}{(\Sigma_{\nu}+1)(\Sigma_{\nu}-1)}\,; \qquad y_1-y_2=-\nu_{(1,2)}\,,\qquad y_4-y_1=\tfrac{1}{2}-\nu_{(4,1)}\,, \\
y_3-y_4 = \tfrac{4\tilde{\mu}(\Sigma_{\nu}-2 \nu^-_{(3,4)})-x[2q_1(\Sigma_{\nu}-2 \nu^-_{(3,4)})+(q_2(1+2 \nu^-_{(3,4)})-q_3(1+2 \Sigma_{\nu}-2 \nu^-_{(3,4)}))(\Sigma_{\nu}-1)+2q\cdot \alpha(\nu^-)(\Sigma_{\nu} -2 \nu^-_{(3,4)})]}{16\tilde{\mu}  -8x q_1+ 4 x (q_2+q_3)(\Sigma_{\nu}-1)-8x q\cdot \alpha(\nu^-)}\,; \\
Y^+_{(1;2)} = \tfrac{4 \pi \tilde{\mu} + 2\pi x[q_1(\Sigma_{\nu}-2)-q\cdot\alpha(\nu^-)]}{\Sigma_{\nu}-1}\,,\qquad
Y^-_{(4;1)} = \tfrac{-4 \pi \tilde{\mu} + 2\pi x[(q_1+q_2+q_3)(\Sigma_{\nu}+2)+q\cdot\alpha(\nu^-)]}{\Sigma_{\nu}+1}\,\cdot
}

\paragraph{Region $\bm{4^-}$:} $ -\tfrac{2\tilde{\mu}}{q_1+q_2 +q_2\Sigma_{\nu}-q\cdot\alpha(\nu^-)}\leq x \leq  -\tfrac{2\tilde{\mu}}{q_1+q_2  +q_3\Sigma_{\nu}-q\cdot\alpha(\nu^-)}$
\eqscn{\rho (x) = \tfrac{-8\tilde{\mu} -4x[ q_1+q_2 (1+\Sigma_{\nu})- q\cdot\alpha(\nu^-)]  }{\Sigma_{\nu}(\Sigma_{\nu}+1)}\,; \\
y_1-y_2=\tfrac{1}{2}-\nu_{(1,2)}\,,\qquad y_3-y_4=-\nu_{(3,4)}\,,\qquad y_4-y_1=-\nu_{(4,1)}\,; \\
Y^-_{(1;2)} = \tfrac{-4 \pi \tilde{\mu} -2 \pi  x[q_1(\Sigma_{\nu}+2)-q\cdot\alpha(\nu^-)]}{\Sigma_{\nu}+1}\,,\qquad Y^+_{(3;4)} = \tfrac{4 \pi \tilde{\mu} + 2\pi x[q_1+q_2+q_3\Sigma_{\nu}-q\cdot\alpha(\nu^-)]}{\Sigma_{\nu}}\,, \\
Y^+_{(4;1)} = \tfrac{4 \pi \tilde{\mu} +2\pi x[q_1+q_2-(q_1+q_2+q_3)\Sigma_{\nu}-q\cdot\alpha(\nu^-)]}{\Sigma_{\nu}}\,\cdot 
}
\paragraph{Region $\bm{4^+}$:} $ \tfrac{2\tilde{\mu}}{ q_1+q_2 -q_2\Sigma_{\nu}+  q\cdot\alpha(\nu^-) }\leq x \leq  \tfrac{2\tilde{\mu}}{ q_1+q_2 -q_3\Sigma_{\nu}+  q\cdot\alpha(\nu^-) }$
\eqscn{\rho (x) = \tfrac{-8\tilde{\mu} +4x [q_1+q_2 - q_3\Sigma_{\nu}+q\cdot\alpha(\nu^-)]  }{\Sigma_{\nu}(\Sigma_{\nu}+1)}\,; \\
y_1-y_2=-\nu_{(1,2)}\,, \qquad y_2-y_3=-\nu_{(2,3)}\,,\qquad y_4-y_1=\tfrac{1}{2}-\nu_{(4,1)}\,; \\
Y^+_{(1;2)} = \tfrac{4\pi\tilde{\mu} +2\pi x[q_1\Sigma_{\nu}-(q_1+q_2)-q\cdot\alpha(\nu^-)]}{\Sigma_{\nu}}\,,\qquad Y^+_{(2;3)} = \tfrac{4\pi\tilde{\mu} -2\pi x[q_1-q_2(\Sigma_{\nu}-1)+q\cdot\alpha(\nu^-)]}{\Sigma_{\nu}} \,, \\
Y^-_{(4;1)} = \tfrac{-4\pi\tilde{\mu} +2\pi x[(q_1+q_2+q_3)(\Sigma_{\nu}+2)+q\cdot\alpha(\nu^-)]}{\Sigma_{\nu }+ 1}\,\cdot
}

Let us visualize all these expressions for $ρ(x)$ and $y(x)$'s in Figure \ref{figA3rhoys} using the numerical values: $q_1=78$, $q_2=2$, $q_3=-29$ and all $Δ$'s equal to $\frac{1}{5}$.
\fig{h!}{\includegraphics[scale=1]{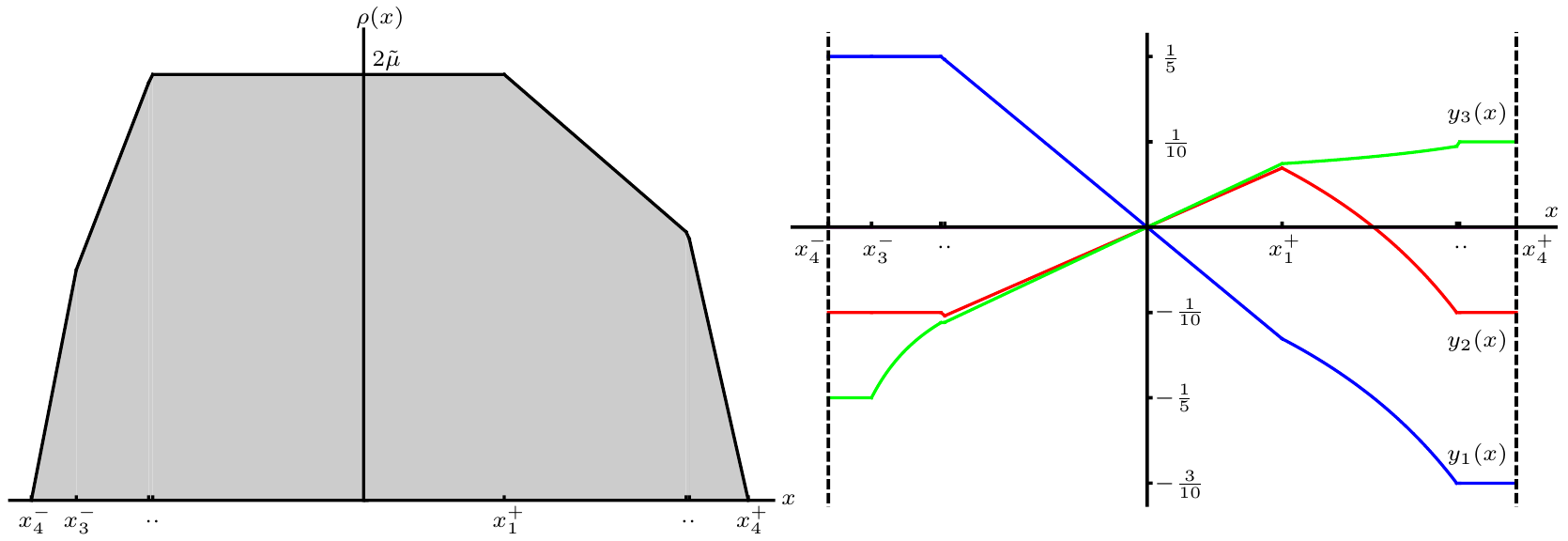}
\caption{Eigenvalue density $ρ(x)$ and distributions $y_a(x)$ for $\wh{A}_3$ quiver ($y_4(x)=0$).}
\label{figA3rhoys}}
With the $ρ(x)$ known in all the regions, we can just use the normalization condition $∫dx ρ(x)=1$ to get $\tilde{μ}$, which gives the Bethe potential $\V\propto \tilde{μ}$. Next, we plot the divergent contributions in Figure \ref{figA3Yss}, which are crucial to get the correct twisted index. Note that all the $Y^±(x)$'s are in the upper half plane as required by consistency.\\
\fig{h!}{\includegraphics[scale=1]{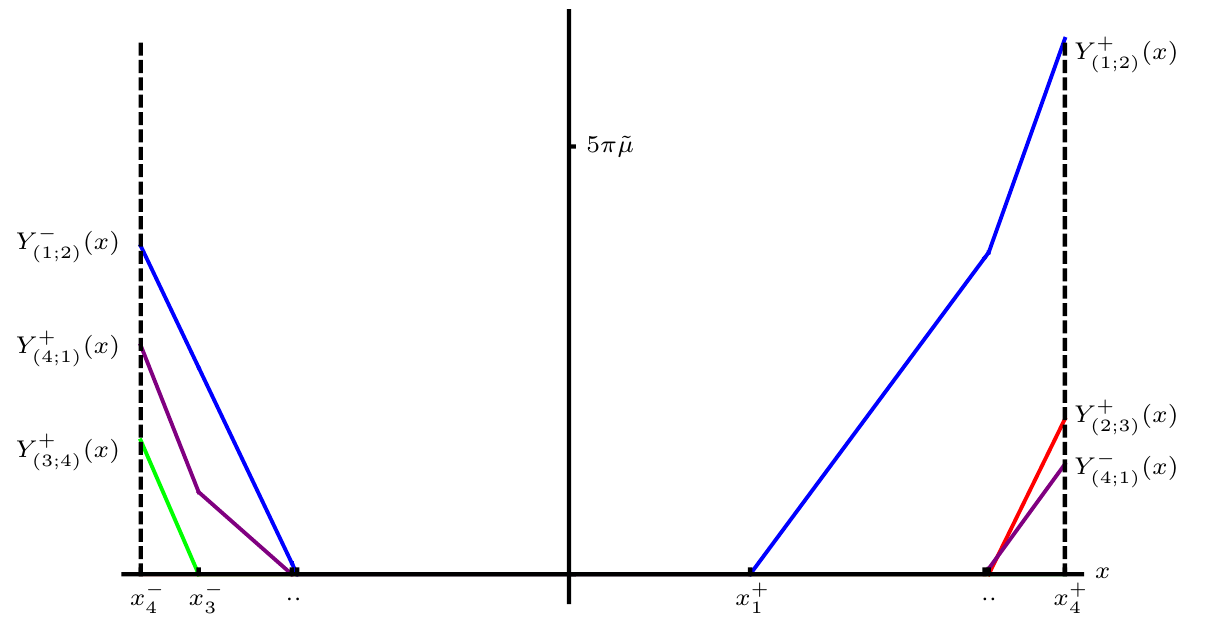}
\caption{Divergent contributions $Y^±_{(a;b)}(x)$ for $\wh{A}_3$ quiver.}
\label{figA3Yss}}

Finally, we have to integrate the expression given in \eqref{indexpOS} with all the $\big\{ρ(x), y_{a,I}(x), Y^±_{(a,I;b,J)}\big\}$ obtained here in each region carefully. The result is a huge expression and unless we take the help of \eqref{genind}, it is hard to make sense of it. Though, we can make sure that the integrated expression and the one obtained via \eqref{genind} are identical, which we have done for both $\wh{A}_2$ and $\wh{A}_3$ to check that \eqref{expYs} does give the correct $Y^±(x)$'s. Thus, instead of writing the full expression for $\wh{A}_3$ here, we present the explicit general result for $\wh{A}_m$ quiver directly.

\subsection[\texorpdfstring{$\wh{A}_m$ Result}{Am Result}]{$\bm{\wh{A}_m}$ Result}
Having discussed the non-trivial case of $\wh{A}_3$ quiver of this class explicitly, we write down the generalization of (well-known) $\wh{A}_1$ and (above-mentioned) $\wh{A}_3$ results quite straightforwardly:
\equ{\V=\frac{4πN^{\sfrac{3}{2}}}{3}\tilde{μ}\quad\text{ with }\quad\frac{1}{\tilde{μ}^2}=32∑_{a=1}^{m+1}\left[\frac{|γ_{a,a+1}|}{\s_a\s_{a+1}}+\frac{|γ_{a,a+1}|}{\s_{a+m+1}\s_{a+m+2}}\right],
\label{genbpvA}}
where only the $\s$'s definitions slightly changes compared to \eqref{bgsdefA}
\equ{\s_a=∑_{b=1}^{m+1}\left[|γ_{a,b}|+2γ_{b,a}ν_{(b,b+1)}^-\right]\,,\quad \s_{a+m+1}=∑_{b=1}^{m+1}\left[|γ_{a,b}| -2γ_{b,a}ν_{(b,b+1)}^-\right].
}
Thus, we see that if we substitute $Δ→2ν$ in \eqref{genvolA}, we get for $\wh{A}$ quivers:
\equ{\frac{1}{μ[2ν]^2}=\frac{1}{16\tilde{μ}[ν]^2} \qquad ⇒ \qquad 4\V[ν]=F_{S^3}[2ν]
}
as promised earlier.

The index is implicitly given by the relation \eqref{genind} but massaging it a little bit, we can give an explicit expression in terms of $\tilde{μ}$ that facilitates checking with the expression given by the integral in \eqref{indexpOS}:
\equ{\bar{\I}=(\fg-1)\frac{4πN^{\sfrac{3}{2}}}{3}\tilde{μ}^3\bigg[\frac{4}{\tilde{μ}^2} -\frac{1}{2}{∑_I}'\(\fn_I -2ν_I\)\frac{∂\big(\frac{1}{\tilde{μ}^2}\big)}{∂ν_I}\bigg].
\label{genindA}
}
The derivative term reads explicitly (after some tedious algebra) as follows:
\eqstn{{∑_I}'\(\fn_I-ν_I\)\frac{∂\big(\frac{1}{\tilde{μ}^2}\big)}{∂ν_I} = -64∑_{a=1}^{m+1}|γ_{a,a+1}|\bigg[\frac{\s_a\(f_{a+1}(\fn) -2f_{a+1}(ν)\)+\(f_a(\fn) -2f_a(ν)\)\s_{a+1}}{\s_a^2\s_{a+1}^2} \nn
+\frac{\s_{a+m+1}\(f_{a+m+2}(\fn) -2f_{a+m+2}(ν)\)+\(f_{a+m+1}(\fn) -2f_{a+m+1}(ν)\)\s_{a+m+2}}{\s_{a+m+1}^2\s_{a+m+2}^2}\bigg],
}
where $f_{a(+m+1)}(\fn)=(-)∑_{b=1}^{m+1}2γ_{b,a}\fn_{(b,b+1)}$, and similarly for $f_a(ν)$.\\

\noindent Let us move on to the $\wh{D}$ quivers now with an explicit solution for $\wh{D}_4$ quiver first.

\subsection[\texorpdfstring{$\wh{D}_4$ Solved}{D₄ Solved}]{$\bm{\wh{D}_4}$ Solved}\label{sec:D4s}
Given the scalings of Table \ref{tabScal}, the boundary $x$-values, $ρ(x)$ and $y_{a,I}(x)$ follow straightforwardly from Section \ref{D4freeEn} so we do not repeat them here. Only the divergent contributions $Y^±(x)$'s are new and we enumerate them region-wise below (we again suppress the $I=1$ index for $a=1,⋯,4$ and $J=2$ for $b=5$). It turns out that there are no kinks in $Y^±(x)$'s here so we write only the new ones appearing in each given region.

\paragraph{Region 1:} $-\frac{2\tilde{\mu}}{2(p_1 +p_2) -2p·α(ν^-)}\leq x \leq \tfrac{2\tilde{\mu}}{2(p_1 +p_2) +2p·α(ν^-)}$
\equn{\text{No } Y^±(x) \text{'s yet.}
}

\paragraph{Region $\bm{2^-}$:} $-\tfrac{2\tilde{\mu}}{2p_1 -2p·α(ν^-)}\leq x \leq - \tfrac{2\tilde{\mu}}{2(p_1 +p_2) -2p·α(ν^-)}$
\equn{Y^-_{(1;5)}=-4\pi\big(\tilde{\mu} +x(p_1 +p_2) -x p·α(ν^-)\big).
}
\paragraph{Region $\bm{2^+}$:} $\tfrac{2\tilde{\mu}}{2(p_1 +p_2) +2p·α(ν^-)}\leq x \leq \tfrac{2\tilde{\mu}}{2p_1 +2p·α(ν^-)}$
\equn{Y^+_{(1;5)}=-4\pi\big(\tilde{\mu} -x(p_1 +p_2) -x p·α(ν^-)\big).
}

\paragraph{Region $\bm{3^-}$:} $-\tfrac{2\tilde{\mu}}{p_1 + p_2 + p_3 + p_4 -2p·α(ν^-)}\leq x \leq  - \tfrac{2\tilde{\mu}}{2p_1 -2p·α(ν^-)}$
\equn{Y^+_{(2;5)}=-4\pi\big(\tilde{\mu} +x p_1 -x p·α(ν^-)\big).
}
\paragraph{Region $\bm{3^+}$:} $\tfrac{2\tilde{\mu}}{2p_1 +2p·α(ν^-)}\leq x \leq \tfrac{2\tilde{\mu}}{p_1+p_2+p_3+p_4 +2p·α(ν^-)}$
\equn{Y^-_{(2;5)}=-4\pi\big(\tilde{\mu} -x p_1 - x p·α(ν^-)\big).
}

\paragraph{Region $\bm{4^-}$:} $-\tfrac{2\tilde{\mu}}{p_1+p_2+p_3-p_4 -2p·α(ν^-)} \leq x \leq - \tfrac{2\tilde{\mu}}{p_1+p_2+p_3+p_4 -2p·α(ν^-)}$
\equn{Y^+_{(4;5)}=-2\pi\big(2\tilde{\mu} +x(p_1 +p_2 +p_3 +p_4) -2x p·α(ν^-)\big).
}
\paragraph{Region $\bm{4^+}$:} $\tfrac{2\tilde{\mu}}{p_1+p_2+p_3+p_4 +2p·α(ν^-)} \leq x \leq \tfrac{2\tilde{\mu}}{p_1+p_2+p_3 -p_4 +2p·α(ν^-)}$
\equn{Y^-_{(4;5)}=-2\pi\big(2\tilde{\mu} -x(p_1 + p_2+p_3 + p_4) -2x p·α(ν^-)\big).
}

\paragraph{Region $\bm{5^-}$:} $-\tfrac{2\tilde{\mu}}{p_1 + p_2 -2p·α(ν^-)} \leq x \leq -\tfrac{2\tilde{\mu}}{p_1+p_2+p_3 -p_4 -2p·α(ν^-)}$
\equn{Y^+_{(3;5)}=-2\pi\big(2\tilde{\mu} +x(p_1 +p_2 +p_3 -p_4) -2x p·α(ν^-)\big).
}
\paragraph{Region $\bm{5^+}$:} $\tfrac{2\tilde{\mu}}{p_1+p_2+p_3 -p_4 +2p·α(ν^-)} \leq x \leq \tfrac{2\tilde{\mu}}{p_1+p_2 +2p·α(ν^-)}$
\equn{Y^-_{(3;5)}=-2\pi\big(2\tilde{\mu} -x(p_1 +p_2 +p_3 -p_4) -2x p·α(ν^-)\big).
}
These $Y^±(x)$'s are plotted in Figure \ref{figD4div} using the numerical values: $p_1=15$, $p_2=8$, $p_3=4$, $p_4=1$ and all $ν$'s equal to $\frac{1}{3}$ and we see that all of them are in the upper half plane as expected.
\fig{h!}{\includegraphics[scale=1]{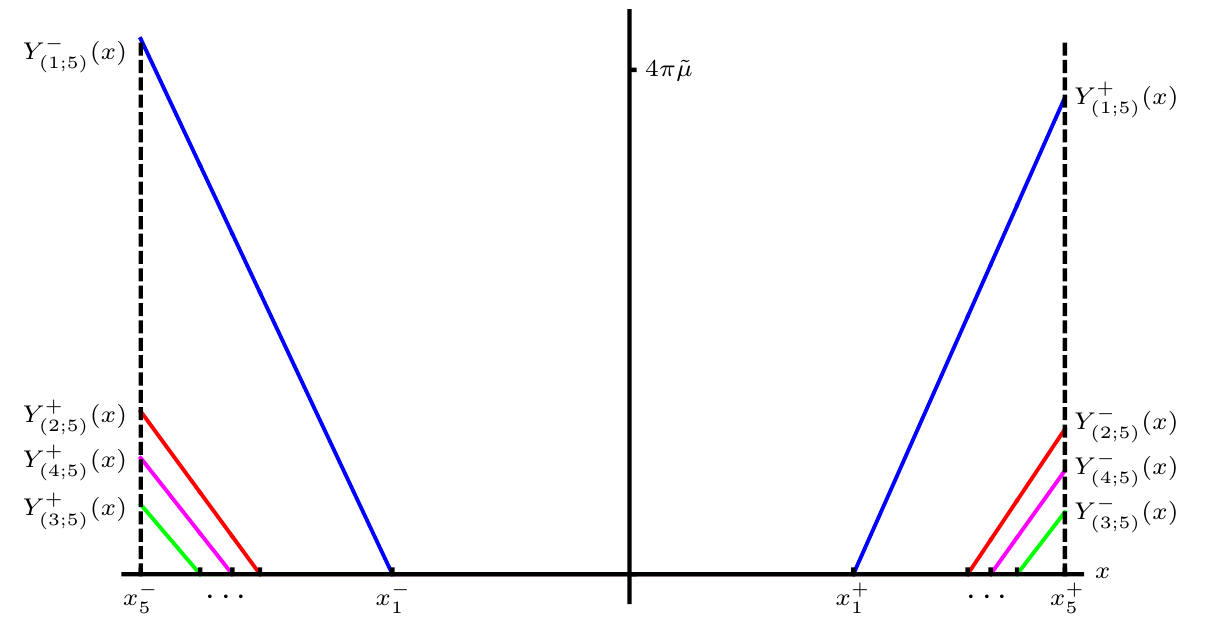}
\caption{Divergent contributions $Y^±_{(a,1;5,2)}(x)$ for $\wh{D}_4$ quiver.}
\label{figD4div}}

Finally, we integrate the expression given in \eqref{indexpOS} by substituting the $\{ρ(x), y_{a,I}(x), Y^±_{(a,I;b,J)}\}$ in each region carefully. The result is again a huge expression and we take help of \eqref{genind} to write it concisely. Before we do that, a comment about insufficiency of \eqref{expYs} for $\wh{D}_n$ with $n>4$ is in order, after which, we will present the explicit general result for $\wh{D}_n$ quiver.

\paragraph{Comment.} Note that the $Y^±(x)$-functions of $\wh{D}_4$ are same in all the regions and no discontinuity appears unlike the case of $\wh{A}_3$. This might lead one to think that all $\wh{D}$ quivers exhibit such a simple behaviour but this is a highly misleading behaviour of $\wh{D}_4$ and does not generalize any further. A true general behaviour appears with $\wh{D}_5$ with discontinuities and subtleties, which breaks down the algebraic system of equations \eqref{expYs} used to solve for $Y^±(x)$'s. This happens because (assuming similar order of $p$'s and similar progression of regions) one of the original inequalities, say, $0<y_{2,1}-y_{5,1}+ν_{(2,5)}<1$ can change to (a subset of) an already existing inequality, say, $-\half<y_{5,1}-y_{6,2}-ν_{(6,5)}<0$ in some region (via saturation sequence like $y_{2,1}→y_{5,2}→y_{6,2}$) and when saturation occurs $y_{5,1}-y_{6,2}-ν_{(6,5)}=0$ in some later region, we will have one more $Y^±(x)$'s to solve than there are equations in \eqref{expYs}! This problem can be solved by demanding the two relevant $Y^±(x)$'s ($Y^-_{(2,1;5,1)}$ and $Y^+_{(5,1;6,2)}$ in above scenario) to be the same. Thus, the algorithm discussed before subsection \ref{sec:A3s} needs to be modified by augmenting \eqref{expYs} with extra equality constraints among $Y^±(x)$'s that appear due to saturation of multiple original inequalities. We have used this modified algorithm for solving $\wh{D}_5$ and $\wh{D}_6$ matrix models and checked that the twisted index obtained from the integral expression \eqref{indexpOS} and that obtained via \eqref{genind} are indeed identical.

\subsection[\texorpdfstring{$\wh{D}_n$ Result}{Dn Result}]{$\bm{\wh{D}_n}$ Result}
It should be no surprise that the result for $\wh{D}$ quivers will look similar to that for $\wh{A}$ quivers:
\equ{\V=\frac{4πN^{\sfrac{3}{2}}}{3}\tilde{μ}\quad\text{ with }\quad\frac{1}{\tilde{μ}^2}=32∑_{a=0}^n\left[\frac{|γ_{a,a+1}|}{\bar{\s}^+_a\bar{\s}^+_{a+1}}+\frac{|γ_{a,a+1}|}{\bar{\s}^-_{a}\bar{\s}^-_{a+1}}\right],
\label{genbpvD}}
where only the $\bar{\s}$'s definitions slightly changes compared to \eqref{bgsdefD}
\equ{\bar{\s}^±_a=∑_{b=1}^n\left[|γ_{a,b}| +|γ_{a,-b}| ±2(γ_{a,b} -γ_{a,-b})α_b(ν^-)\right] -4|γ_{a,n+1}|\,.
}
Thus, we again see that upon substituting $Δ→2ν$ in \eqref{genvolD}, we get for $\wh{D}$ quivers
\equ{\frac{1}{μ[2ν]^2}=\frac{1}{16\tilde{μ}[ν]^2} \qquad ⇒ \qquad 4\V[ν]=F_{S^3}[2ν]
}
as expected. One caveat here is that the result for $\wh{A}_m$ quivers is an exact result whereas that for $\wh{D}_n$ quivers is a conjecture. This boils down to the polygon formulas \eqref{coneA} and \eqref{coneD}. While the former is a proven solution to the $\wh{A}_m$ matrix model\cite{Gulotta:2011aa}, the latter is a conjecture that we have checked for $\wh{D}_n$ matrix model up to $n=10$.

Finally, the index can be written explicitly as follows:
\eqsc{\I=(\fg-1)\frac{4πN^{\sfrac{3}{2}}}{3}\tilde{μ}^3\bigg[\frac{4}{\tilde{μ}^2} -\frac{1}{2}{∑_I}'\(\fn_I -2ν_I\)\frac{∂\big(\frac{1}{\tilde{μ}^2}\big)}{∂ν_I}\bigg]\,; \label{genindD}\\
{∑_I}'\(\fn_I-2ν_I\)\frac{∂\big(\frac{1}{\tilde{μ}^2}\big)}{∂ν_I}= -64\(f(\fn)-2f(ν)\) \bigg(\frac{|γ_{-1,1}|\(\bar{\s}^-_1+\bar{\s}^+_1\)}{(\bar{\s}^-_1)^2(\bar{\s}^+_1)^2}+∑_{±,a=1}^n\frac{|γ_{a,a+1}|\(\bar{\s}^±_a+\bar{\s}^±_{a+1}\)}{(\bar{\s}^±_a)^2(\bar{\s}^±_{a+1})^2}\bigg), \nonumber
}
where $f(\fn)=∑_{b=1}^n 2(γ_{a,b} -γ_{a,-b})α_b(\fn)$, and similarly for $f(ν)$. Due to the fact that $(γ_{a,b} -γ_{a,-b})=2p_b$ does not depend on the subscript $a$, these $f(·)$'s become an overall factor and the explicit expression for $\wh{D}$ quivers' index simplifies considerably compared to the analogous expression for $\wh{A}$ quivers.

\section{Summary and Outlook}\label{sec:SO}
This paper contains two interconnected results:
\begin{description}
\item[Volume:] We computed the explicit free energy $F_{S^3}$ for $\wh{D}$ quivers in terms of the R-charges $Δ_{(a,b)}$ of the bifundamentals, obtained by combining \eqref{genextF} and \eqref{genvolD}. According to AdS/CFT correspondence, the formula \eqref{genvolD} provides a prediction for the volumes of certain Sasaki-Einstein 7-manifolds $Y_7$, which describe the AdS${}_4×Y_7$ M-theory duals.
\item[Entropy:] We computed the explicit twisted index $\I$ for $\wh{AD}$ quivers, \eqref{genindA} and \eqref{genindD}, in terms of the chemical potentials $ν_{(a,b)}$ and flavour fluxes $\fn_{(a,b)}$. We expect that the extremization of these formulas with respect to $ν$'s leading to the expression $\I(ν(\fn),\fn)$ reproduces the macroscopic entropy of the dual black hole solutions in the 4d gauged supergravity uplifted to M-theory with the above-mentioned $Y_7$'s. In the simplifying case of universal twist, the extremization procedure is automatic, leading to $\fn_I=2ν_I$ and $S_{BH}=\I=(g-1)F_{S^3}$ follows via the relation \eqref{genind} for $\wh{ADE}$ quivers as shown holographically in \cite{Bobev:2017uzs,Azzurli:2017kxo}.
\end{description}

Along the way, we computed the large $N$ limit of the partition functions for 3d $\N=2$ quiver theories on $S^3$ and $Σ_{\fg}×S^1$ involving bifundamental and fundamental matters. We obtained constraints on relevant parameters ($Δ$ for $F_{S^3}$ and $\{ν,\fn\}$ for $\I$) under the requirement that the resulting matrix model be local, leading to a large class of CSm quiver theories including the $\wh{ADE}$ quivers. The fundamental matters contribute in a trivial way and that contribution can be included in the results presented here following \cite{Crichigno:2012sk,Jain:2015kzj}. An intermediate construction to obtain the twisted index is that of the Bethe potential $\V$, which we find is related to the free energy via $F_{S^3}[2ν]=4\V[ν]$ with an explicit matching of the matrix model. It was shown in \cite{Hosseini:2016tor} that for $\wh{A}$ quivers and related theories, this relation is true off-shell too but with a \emph{different} numerical factor. We extended this result to $\wh{DE}$ quivers and showed that the relation holds true in all the integration regions with the \emph{same} numerical factor of 4. This fact fits nicely with the simpler proof of the relation \eqref{genind} between the twisted index and the Bethe potential provided in the main text.

We note that one could study these theories on more general Seifert manifolds as discussed in \cite{Closset:2017zgf,Toldo:2017qsh,Closset:2018ghr}. The $\M_{\fg,p}$ manifolds include both the manifolds studied here as in $\M_{0,1}=S^3$ and $\M_{\fg,0}=Σ_{\fg}×S^1$. In this framework, the observation $4\V[ν]=F_{S^3}[2ν]$ in the present context may be easily explained following the logic of \cite{Toldo:2017qsh}. In addition, it should be possible to generalize the results presented here straightforwardly to these manifolds.

An Elephant in the room is the fact that expressions for free energies of $\wh{E}_{6,7,8}$ are missing in this paper. As is well-known, even in the $\N=3$ case\cite{Crichigno:2012sk} the known expressions are valid only for a subset of CS levels. An all-encompassing formula in terms of roots or graphs as in the case of $\wh{AD}$ quivers is not known for them. So we refrained from giving the $\N=2$ extensions of the $\N=3$ formulas but comment that it would be much more interesting to figure out the fully general volume formula for $\wh{E}$ quivers. The Fermi-gas formalism\cite{Marino:2011eh,Marino:2012az,Assel:2015hsa,Moriyama:2015jsa} could be a helpful tool in this quest, given that the polygon formula appears naturally as a Fermi surface in this formalism.

Finally, it goes without saying that computing volumes of the Sasaki-Einstein 7-manifolds explicitly and constructing explicit M-theory duals for $\wh{ADE}$ quivers with non-universal flavour fluxes would be an interesting exercise to test the AdS/CFT correspondence.

\section*{\centering Acknowledgements}
DJ thanks P. M. Crichigno for comments, corrections and suggestions on the draft of this paper. AR is indebted to Arnab Kundu for various discussions related to this work specifically and to physics in general. He acknowledges support from the Department of Atomic Energy, Government of India.

\noindent We also acknowledge that most of this work would not have been possible without \texttt{Mathematica v11.3}\cite{Mathematica}.

\appendix
\section{Derivations and Proofs}\label{app:DP}
We collect here some details of the calculations that went into evaluating free energy and twisted index of the CS quiver gauge theories. Our derivations have considerably more overlap with \cite{Jafferis:2011zi,Toldo:2017qsh} than with \cite{Benini:2015eyy,Hosseini:2016tor}.

\subsection[\texorpdfstring{$F_{S^3}$}{Free Energy}]{$\bm{F_{S^3}}$}
We start with the expression whose large $N$ limit is to be obtained:
\eqsn{F_{S^3} &≈ -\i πN∫dx ρ(x)∑_{a,I}k_a \big(N^αx +\i y_{a,I}(x)\big)^2 \\
&-N^2∫dxdx'ρ(x)ρ(x')∑_{a,I,J}\log\big|2\sinh\big(πN^α(x-x')+\i π(y_{a,I}(x) -y_{a,J}(x'))\big)\big| \\
&-N^2∫dxdx'ρ(x)ρ(x')∑_{(a,b)∈E}∑_{I,J}\ell\big(1-Δ_{(a,b)} +\i N^α(x-x') -(y_{a,I}(x) -y_{b,J}(x'))\big) \\
&-N∫dxρ(x)∑_{\mathclap{a,\{f^a\},I}}\ell\big(1-Δ_{f^a} +\i N^αx -y_{a,I}\big)-N∫dxρ(x)∑_{\mathclap{a,\{\bar{f}^a\},I}}\ell\big(1-\bar{Δ}_{f^a} -\i N^αx +y_{a,I}\big).
}
The four lines correspond to four different contributions as follows:

\paragraph{Chern-Simons.} This is pretty straightforward
\eqs{&-\i πN∫dx ρ(x)∑_{a,I}k_a \big(N^{2α}x^2 +2\i N^α x y_{a,I}(x) -y_{a,I}(x)^2\big) \nn
=&-\i π N^{1+2α}∫dx ρ(x)∑_a(n_ak_a) x^2 +2πN^{1+α}∫dx ρ(x)x∑_{a,I}k_ay_{a,I}(x) +\O(N)\,,
}
and we get the first line of equation \eqref{FS3fullint}.

\paragraph{Vectors.} This is again straightforward following \cite{Gulotta:2011vp}. We change variables $πN^α(x-x')→ξ$ which implies to leading order $ρ(x')→ρ(x)$ and $y(x')→y(x)$ so we get:
\eqs{&-\frac{N^{2-α}}{π}∫dx\,ρ(x)^2∫_{-M}^M dξ ∑_{a,I,J}\log\big[2\sinh\big(|ξ|+\i π\sgn(ξ)δy_{a,IJ}(x)\big)\big] \nn
=&-\frac{N^{2-α}}{π}∫dx ρ(x)^2\bigg[∑_an_a^2\bigg(M^2 +\frac{π^2}{12}\bigg) -∑_{a,I,J}\frac{1}{4}\arg\(e^{2π\i(δy_{a,IJ}(x)-\sfrac{1}{2})}\)^2\bigg]\,.
\label{vecont}}
The $\arg()$ term appears in second line of \eqref{FS3fullint} and requiring the value of exponent to lie in the principal branch (i.e., $-\half<δy_{a,IJ}-\half<\half$) gives the relevant inequality of \eqref{FS3ineq}. Now we show that the divergent $\O(M^2)$ (and a finite $\O(1)$) piece cancels a similar term coming from matter contributions.

\paragraph{(Anti-)Bifundamentals.} This is slightly tricky but assuming equal number of bifundamental and anti-bifundamental matter at each edge initiates a few cancellations. Following through a direct calculation (refer to \cite{Jafferis:2011zi} for a slightly different derivation) with a similar change of variables $N^α(x-x')→ξ$ as above and defining $\P_{(a,b)}=1-Δ_{(a,b)}-δy_{ab,IJ}(x)$, $\P_{(b,a)}=1-Δ_{(b,a)} +δy_{ab,IJ}(x)$, we get (using the definition \eqref{defell} of $\ell(z)$):
\eqs{&-N^{2-α}∫dxρ(x)^2∫_{-M}^M dξ ∑_{(a,b)∈E}∑_{I,J}\(\ell\big(\P_{(a,b)} +\i ξ\big) +\ell\big(\P_{(b,a)} -\i ξ\big)\) \nn
=&-N^{2-α}∫dxρ(x)^2 ∑_{(a,b)∈E}∑_{I,J}\pmat{-\frac{\i π}{3}M^3 +\frac{\i π}{6}M\(-1+6\P_{(a,b)}^2\) \\ -\frac{\i}{2π}\(M+\i\P_{(a,b)}\)\Li_2\big(e^{2π(M+\i\P_{(a,b)})}\big) \\ +\frac{\i}{2π^2}\Li_3\big(e^{2π(M+\i\P_{(a,b)})}\big) }+\(\P_{(a,b)}→\P_{(b,a)}\).
}
The following identities are required for polylogs when $M→+∞$ and $a∈\bR$ to simplify the above expressions:
\eqs{\Li_1\big(e^{M+2π\i a}\big) &→ -M-\i \arg\big(e^{2π\i(a+\sfrac{1}{2})}\big) \\
\Li_2\big(e^{M+2π\i a}\big) &→ -\tfrac{M^2}{2} -\i M\arg\big(e^{2π\i(a+\sfrac{1}{2})}\big) -\(\tfrac{π^2}{6} -\half\arg\big(e^{2π\i(a+\sfrac{1}{2})}\big)^2\) \\
\Li_3\big(e^{M+2π\i a}\big) &→ -\tfrac{M^3}{6} -\tfrac{\i}{2}M^2\arg\big(e^{2π\i(a+\sfrac{1}{2})}\big) -M\(\tfrac{π^2}{6} -\half\arg\big(e^{2π\i(a+\sfrac{1}{2})}\big)^2\) \nn
&\qquad\quad -\tfrac{\i}{6}\arg\big(e^{2π\i(a+\sfrac{1}{2})}\big)\(π^2-\arg\big(e^{2π\i(a+\sfrac{1}{2})}\big)^2\).
}
The terms at $\O(M^3)$ simply cancel without the need to add $\P_{(a,b)}$ and $\P_{(b,a)}$ contributions. Upon adding both contributions\footnote{We also scaled $M→\frac{M}{π}$ to match the transcendentality of other terms and to compare with \eqref{vecont}.}, we get the following terms at various orders of $M$ (suppressing the overall $∫dxρ(x)^2$):
\eqs{\O(M^2): &-\frac{N^{2-α}}{π}∑_{(a,b)∈E}n_an_b\(-2+Δ_{(a,b)}+Δ_{(b,a)}\); \label{bifuncont}\\
\O(M): &\,\frac{\i N^{2-α}}{4π^2}∑_{(a,b)∈E}∑_{I,J}\(π^2(1-4\P_{(a,b)}^2)+\arg(⋯)\big(4π\P_{(a,b)}-\arg(⋯)\big)\) \nn
&+\(\P_{(a,b)}→\P_{(b,a)}\); \label{om0}\\
\O(1): &-\frac{N^{2-α}}{π}∑_{(a,b)∈E}n_an_b\frac{π^2}{12}\(-2+Δ_{(a,b)}+Δ_{(b,a)}\) \nn
&-\frac{N^{2-α}}{4π}∑_{(a,b)∈E}∑_{I,J}\bigg(\P_{(a,b)}\arg(⋯)^2 +\frac{1}{3π}\arg(⋯)\(π^2-\arg(⋯)^2\)\bigg) \nn
&+\(\P_{(a,b)}→\P_{(b,a)}\),
\label{Rbifuncont}}
where $⋯≡e^{2π\i(\P_{(a,b)}+\sfrac{1}{2})}$. The $\O(M^2)$ term from \eqref{vecont} and \eqref{bifuncont} add up to give the $\wh{ADE}$ constraint \eqref{ADEconditionFS3} that was derived from the saddle point equation in the main text. This kills the $\O(M^2)$ as well as one of the $\O(1)$ contributions from VM and (anti-)bifundamental MM. Again, requiring the exponent inside $\arg(⋯)$ to lie in the principal branch gives the relevant inequalities of \eqref{FS3ineq}. Effectively, we have to impose $0≤δy_{ab,IJ}+Δ_{(a,b)}<1$ and $-1<δy_{ab,IJ}-Δ_{(b,a)}≤0$ (the equalities denote the saturation values at region boundaries) which lead to the $\O(M)$ contribution \eqref{om0} vanishing identically. The remaining $\O(1)$ contribution from \eqref{Rbifuncont} is what appears in the third and fourth lines of \eqref{FS3fullint}.

\paragraph{(Anti-)Fundamentals.} This contribution follows directly from the definition of the $\ell(z)$-function after using the polylog identities given above:
\eqst{-N∫dxρ(x)∑_{a,\{f^a\},I}\(-\frac{\i}{2}N^{2α}πx^2 -N^α π|x|\(1-Δ_{f^a}-y_{a,I}(x)\)+\O(1)\) \\
-N∫dxρ(x)∑_{a,\{\bar{f}^a\},I}\(\frac{\i}{2}N^{2α}πx^2 -N^α π|x|\(1-\bar{Δ}_{f^a}+y_{a,I}(x)\)+\O(1)\),
}
which is what appears in the last two lines of \eqref{FS3fullint}.

\subsection[\texorpdfstring{$\V$}{Bethe Potential}]{$\bm{\V}$}
We again start with the expression whose large $N$ limit is to be obtained:
\eqsn{\V &=-\i∑_{a,i}πk_a(u_a^i)^2 +\frac{1}{2}∑_{a,i,j}π\sgn(j-i)(u_a^i-u_a^j)  -\i∑_{(a,b)∈E}∑_{i,j}\Big(v(u_a^i-u_b^j+\i ν_{(a,b)}) \\
&\quad +v(u_b^j-u_a^i+\i ν_{(b,a)})\Big) -\i∑_{a,i}∑_{\{f^a\}}v(u_a^i+\i ν_{f^a}) -\i∑_{a,i}∑_{\{\bar{f}^a\}}v(-u_a^i+\i \bar{ν}_{f^a}) \\
\V&≈-\i π N∫dxρ(x)∑_{a,I}k_a\(N^α x+\i y_{a,I}(x)\)^2 \\
&+\frac{π}{2}N^2∫dxdx'ρ(x)ρ(x')∑_{a,I,J}\sgn(x'-x)\(N^α(x-x')+\i δy_{a,IJ}(x)\) \\
&-\i N^2∫dxdx'ρ(x)ρ(x')∑_{(a,b)∈E}∑_{I,J}\Big(v\big(N^α(x-x')+\i(δy_{ab,IJ}(x)+ν_{(a,b)})\big) +(ν_{(b,a)}\text{ term})\Big) \\
&-\i N∫dxρ(x)∑_{\mathclap{a,\{f^a\},I}}v\(N^α x+\i (y_{a,I}(x)+ν_{f^a})\) -\i N∫dxρ(x)∑_{\mathclap{a,\{\bar{f}^a\},I}}v\(-N^α x-\i (⋯)\).
}
The four lines above correspond to different contributions as follows:

\paragraph{Chern-Simons.} This is straightforward to compute and after imposing $n_ak_a=0$ leads to just one term at leading order:
\equ{πN^{1+α}∫dxρ(x)2x∑_{a,I}k_ay_{a,I}(x),
}
which appears as the first term in \eqref{BetheV}.

\paragraph{Vectors.} This integral is to be done using the same transformation $(N^α(x-x')→ξ)$ as was done in the case of $F_{S^3}$ which gives us:
\equ{\frac{π}{2}N^{2-α}∫dxρ(x)^2∫_{-M}^M dξ ∑_{a,I,J}\sgn(-ξ)\(ξ+\i δy_{a,IJ}(x)\) =-\frac{π}{2}N^{2-α}M^2∑_an_a^2∫dx ρ(x)^2\,,
\label{Vvecont}}
which is clearly divergent but we expect this to be cancelled by the bifundamental matter contributions as we show next.

\paragraph{(Anti-)Bifundamentals.} As done in the $F_{S^3}$ case, we change variables $N^α(x-x')→ξ$ and define $\P_{(a,b)}=δy_{ab,IJ}(x)+ν_{(a,b)}$, $\P_{(b,a)}=-δy_{ab,IJ}(x)+ν_{(b,a)}$ to get (using the definition \eqref{defvee} of $v(z)$):
\eqs{&-\i N^{2-α}∫dxρ(x)^2∫_{-M}^M dξ ∑_{(a,b)∈E}∑_{I,J}\(v\big(ξ+\i\P_{(a,b)}\big) +v\big(-ξ+\i\P_{(b,a)}\big)\) \nn
=&-\i N^{2-α}∫dxρ(x)^2 ∑_{(a,b)∈E}∑_{I,J}\pmat{\frac{π}{3}M^3 -\frac{π}{6}M\(1+6\P_{(a,b)}^2\) \\ +\frac{1}{4π^2}\Li_3\big(e^{2π(M+\i\P_{(a,b)})}\big)}+\(\P_{(a,b)}→\P_{(b,a)}\).
}
Again, the terms at $\O(M^3)$ simply cancel after using the polylog identities and we get the following terms at various orders of $M$ (suppressing $∫dxρ(x)^2$):
\eqs{\O(M^2): &-N^{2-α}π∑_{(a,b)∈E}n_an_b\(-1+ν_{(a,b)}+ν_{(b,a)}\); \label{Vbifuncont}\\
\O(M): &\frac{\i N^{2-α}}{4π}∑_{(a,b)∈E}∑_{I,J}\(π^2(1+4\P_{(a,b)}^2)+\arg(⋯)^2\)+\(\P_{(a,b)}→\P_{(b,a)}\); \label{Vom0}\\
\O(1): &-\frac{N^{2-α}}{24π^2}∑_{(a,b)∈E}∑_{I,J}\Big(\arg(⋯)\(π^2-\arg(⋯)^2\)\Big) +\(\P_{(a,b)}→\P_{(b,a)}\),
\label{VRbifuncont}}
where $⋯≡e^{2π\i(\P_{(a,b)}+\sfrac{1}{2})}$. The $\O(M^2)$ term from \eqref{Vvecont} and \eqref{Vbifuncont} can be cancelled now leading to the $\wh{ADE}$ constraint of \eqref{ADEconditionTTI}. The $\O(M)$ term leads to an imaginary constant in $\V$ which does not affect the BAEs so we ignore it. The $\O(1)$ contribution in \eqref{VRbifuncont} is what appears in \eqref{BetheV}.

\paragraph{(Anti-)Fundamentals.} This contribution follows directly from the definition of the $v(z)$-function after using the polylog identities as done in the case of $F_{S^3}$:
\eqst{-\i N∫dxρ(x)∑_{a,\{f^a\},I}\(-\half N^{2α}πx^2 -\i N^α π|x|\(-1+y_{a,I}(x)+ν_{f^a}\)+\O(1)\) \\
-\i N∫dxρ(x)∑_{a,\{\bar{f}^a\},I}\(\half N^{2α}πx^2 +\i N^α π|x|\(y_{a,I}(x)-\bar{ν}_{f^a}\)+\O(1)\).
}
After using the simplifying condition $f^a=\bar{f}^a$, $x^2$ and $y(x)$ terms above cancel and we get $(n_F-ν_F)$ term of \eqref{BetheV}.

\paragraph{``Homogeneous'' $\bm{\V}$.} $\V$ as written in the main text is not homogeneous with respect to $ν$'s so first of all, we make $\V$ scale with $ν$'s (at the very least) by defining new shifted $ν$'s or multiplying $2ν^+$ to $γ_{a,b}$ terms in $\bar{\s}$'s. Next, let us look at the on-shell expression for Bethe potential of $\wh{D}_4$ quiver and combine terms over a common denominator. We find that the highest power of $ν$ that survives (after a few simplifications) in the numerator is $6$ whereas that in the denominator is $10$. The difference between them is of course $-4$. Now, looking at the general expression for $\V$ in \eqref{genbpvD}, we note that a pair of terms with $\bar{\s}_5^±$ get included when moving up to $\wh{D}_5$, which changes the degree of both numerator and denominator by 2 so the difference still continues to be $-4$. Thus, by induction, we get that the highest power of $ν$ in the numerator of $\V$ for $\wh{D}_n$ is $2(n-1)$ whereas that in the denominator is $2(n+1)$. The difference between them is of course $-4$ meaning that $\frac{1}{μ^2}$ is of degree $-4$, which translates to the fact that $\V\propto μ$ is of degree $2$ as used in the main text. The same counting works for $\wh{A}$ quivers (starting with $\wh{A}_1$ which has $\V$ of obvious degree $-4$) and should work out for $\wh{E}$ quivers too.

\subsection[\texorpdfstring{$\I$}{Twisted Index}]{$\bm{\I}$}
The expression to be massaged at large $N$ is obtained from \eqref{ZTIbaeexp} (up to a factor of $(\fg-1)$):
\eqsn{\I &=∑_{\mathclap{a,i,j≠i}}v'(u_a^i-u_a^j) +∑_{(a,b)∈E}∑_{i,j}\Big(\(\fn_{(a,b)}-1\)v'(u_a^i-u_b^j+\i ν_{(a,b)}) +\(\fn_{(b,a)}-1\) \\
& ×v'(u_b^j-u_a^i+\i ν_{(b,a)})\Big) +∑_{\mathclap{a,\{f^a\},i}}\(\fn_{f^a}-1\)v'(u_a^i+\i ν_{f^a}) +∑_{\mathclap{a,\{\bar{f}^a\},i}}\(\bar{\fn}_{f^a}-1\)v'(-u_a^i+\i \bar{ν}_{f^a}) +\log\B \\
\I &≈N^2∫dxdx'ρ(x)ρ(x')∑_{a,I}v'\big(\sqrt{N}(x-x')+\i δy_{a,IJ}(x)\big) \\
&+N^2∫dxdx'ρ(x)ρ(x')∑_{(a,b)∈E}∑_{I,J}\Big(\!\!\(\fn_{(a,b)}-1\)v'\big(\sqrt{N}(x-x')+\i \P_{(a,b)}\big) +(\fn_{(b,a)} \text{ term})\Big) \\
&+N∫dxρ(x)∑_{\mathclap{a,\{f^a\},I}}\(\fn_{f^a}-1\)v'\big(\sqrt{N} x+\i(y_{a,I}(x)+ν_{f^a})\big) +N∫dxρ(x)∑_{\mathclap{a,\{\bar{f}^a\},I}}\(\bar{\fn}_{f^a}-1\)v'(⋯) \\
&+N∫dx ρ(x) \big[\O(\sqrt{N}) \text{ term from }\log\B\big]\,.
}
As before, we work out serially each of the four lines above corresponding to different contributions as follows:

\paragraph{Vectors.} The function $v'(α(u))$ appears as the vector contribution because $∏_α e^{πα(u)}=1$. Using the polylog identities, we simply get:
\equ{-N^{\sfrac{3}{2}}\frac{1}{2π}∫dx ρ(x)^2\bigg[∑_an_a^2\bigg(\frac{M^2}{2} +\frac{π^2}{6}\bigg) -∑_{a,I,J}\frac{1}{2}\arg\(e^{2π\i(δy_{a,IJ}(x)-\sfrac{1}{2})}\)^2\bigg]\,.
\label{vecindcont}}
The $\arg()$ term appears in \eqref{indexpOS} and again the rest of the terms will cancel the next contribution.\footnote{We do not explicitly write the imaginary $\O(M)$ terms but one can check that using $Σ_{a,I,J}δy_{a,IJ}(x)=0$, $y(x)$-dependence does not survive and using the $\wh{ADE}$ constraint, constant pieces cancel between vector and bifundamental contributions, just like the $\(\frac{M^2}{2}+\frac{π^2}{6}\)$ terms.}

\paragraph{(Anti-)Bifundamentals.} The evaluation of this term is same as above and we get:
\eqst{-N^{\sfrac{3}{2}}\frac{1}{2π}∫dx ρ(x)^2∑_{\mathclap{(a,b)∈E}}\,\bigg\{\(\fn_{(a,b)}-1\)\bigg[n_an_b\bigg(\frac{M^2}{2} +\frac{π^2}{6}\bigg) -∑_{I,J}\frac{1}{2}\arg\(e^{2π\i(δy_{ab,IJ}(x)+ν_{(a,b)}-\sfrac{1}{2})}\)^2\bigg] \\
+\(\fn_{(b,a)}-1\)\bigg[n_an_b\bigg(\frac{M^2}{2} +\frac{π^2}{6}\bigg) -∑_{I,J}\frac{1}{2}\arg\(e^{2π\i(-δy_{ab,IJ}(x)+ν_{(b,a)}-\sfrac{1}{2})}\)^2\bigg]\bigg\}\,.
}
We see that the $\O(M^2)$ (and the $\frac{π^2}{6}$) term cancels between the above expression and \eqref{vecindcont} if
\equ{∑_an_a^2+∑_{(a,b)∈E}\(\fn_{(a,b)}+\fn_{(b,a)}-2\)n_an_b=0\,,
}
which is same as \eqref{FFxcondition}, leading to the $\wh{ADE}$ constraint when $\fn_{(a,b)}+\fn_{(b,a)}=1$ as mentioned in the main text. The remaining $\arg()$ terms upon imposing the constraint on $ν$'s and $\fn$'s gives the finite bifundamental contributions to \eqref{indexpOS}.

In addition, there are contributions from bifundamentals when $v'(z)$ diverges at $z=0$, which are called tail contributions in the literature\footnote{This terminology makes sense in the case of ABJM and other theories which have only 2 regions. As we saw in explicit examples of $\wh{A}_3$ and $\wh{D}_4$, there is no clear demarcation between tail regions and non-tail regions.}. This happens when $x=x'$ {\bf and} $δy_{ab,IJ}(x)+ν_{(a,b)}=0$ or $δy_{ab,IJ}(x)-ν_{(b,a)}=0$. Thus, following the analysis around \eqref{expYs}, we can write for the divergent bifundamental contributions to $\I$ (for specific $a,b,I,J$ values):
\eqs{δy_{ab,IJ}(x)+ν_{(a,b)}=0 &: -N∫dx ρ(x)\fn_{(b,a)}v'\(e^{-N^{\sfrac{1}{2}}Y^+_{(a,I;b,J)}(x)}\) ≈ -N^{\sfrac{3}{2}}∫dx ρ(x) \fn_{(b,a)} Y^+_{(a,I;b,J)}(x) \nn
δy_{ab,IJ}(x)-ν_{(b,a)}=0 &: -N∫dx ρ(x)\fn_{(a,b)}v'\(e^{-N^{\sfrac{1}{2}}Y^-_{(a,I;b,J)}(x)}\) ≈ -N^{\sfrac{3}{2}}∫dx ρ(x) \fn_{(a,b)} Y^-_{(a,I;b,J)}(x).
\label{bifunsatYs}}

\paragraph{(Anti-)Fundamentals.} This is again a straightforward computation using the polylog identities and imposing $f^a=\bar{f}^a$ leads to
\equ{-N^{\sfrac{3}{2}}∫dxρ(x) π|x|∑_{a,\{f^a\},I}\(\fn_{f^a}+\bar{\fn}_{f^a}-2\),
}
which is what appears in the last term of \eqref{indexpOS}.

\paragraph{Hessian.} The $\log\B$ term is naïvely of $\O(N\log N)$ but due to the diverging nature of $v''(z)=-π\coth(πz)$\footnote{Recall that $\V$ depends on $v(u_a^i -u_b^j+⋯)$ and $\B=\det_{ai,bj}\frac{∂^2\V}{∂u_a^i∂u_b^j}$ so $\B$ would depend on $v''(⋯)$.} at $z=0$, there arise terms of $\O(\sqrt{N})$ that contribute to $\I$ at the leading order. A careful splitting of divergent and non-divergent terms of $\log\B$ has been discussed in \cite{Benini:2015eyy}. We will just show how the divergent term contributes to the large $N$ limit of the index. The divergence in $\B$ occurs exactly when the bifundamental contribution diverges so the Hessian in large $N$ limit contributes:
\eqsn{δy_{ab,IJ}(x)+ν_{(a,b)}=0 &: N∫dx ρ(x) \log v''\(e^{-N^{\sfrac{1}{2}}Y^+_{(a,I;b,J)}(x)}\) ≈ +N^{\sfrac{3}{2}}∫dx ρ(x) Y^+_{(a,I;b,J)}(x) \nn
δy_{ab,IJ}(x)-ν_{(b,a)}=0 &: N∫dx ρ(x) \log v''\(e^{-N^{\sfrac{1}{2}}Y^-_{(a,I;b,J)}(x)}\) ≈ +N^{\sfrac{3}{2}}∫dx ρ(x) Y^-_{(a,I;b,J)}(x)\,,
}
where we used that $\lim_{z→0}v''(z)=-z^{-1}$. Combining above expressions with the bifundamental contributions \eqref{bifunsatYs}, we get the total divergent contribution to the twisted index as:
\equ{δy_{ab,IJ}(x)±ν_{(·,·)}=0 : N^{\sfrac{3}{2}}∫dx ρ(x) \fn_{(·,·)} Y^±_{(a,I;b,J)}(x)\,,
}
which appears in the last line of \eqref{indexpOS} with the Kronecker $δ$ enforcing the condition on $δy(x)$'s.

\newpage
\references{bib3d} 

\end{document}